\documentclass[a4paper,fleqn]{cas-dc}
\usepackage{enumitem}
\usepackage[authoryear]{natbib}
\usepackage[figuresright]{rotating}
\usepackage{moreverb}
\usepackage{amsmath,amssymb,amsfonts}
\usepackage{algorithmic}
\usepackage{url}
\usepackage{graphicx}
\usepackage{textcomp}
\usepackage{xcolor}
\usepackage{tcolorbox}
\usepackage{ragged2e}
\usepackage{balance}
\usepackage{colortbl}
\usepackage{multirow}
\usepackage{listings}
\usepackage{color}

\newcommand{\black}{\textcolor[rgb]{0.00,0.00,0.00}}
\usepackage{CJKutf8}
\usepackage{siunitx}
\usepackage{rotfloat}
\usepackage{float}
\usepackage{lscape}
\usepackage{amssymb,mathrsfs,amsmath}
\usepackage{adjustbox}
\usepackage{threeparttable}
\usepackage{hyperref}
\usepackage{array}
\usepackage{subcaption}    

\newcolumntype{M}[1]{>{\arraybackslash}m{#1}}

\definecolor{icolor}{RGB}{255, 0, 0}
\definecolor{ihcolor}{RGB}{0, 255, 0}
\definecolor{mcolor}{RGB}{255, 255, 0}
\definecolor{ncolor}{RGB}{0, 0, 255}
\definecolor{ecolor}{RGB}{255, 165, 0}

\def\tsc#1{\csdef{#1}{\textsc{\lowercase{#1}}\xspace}}
\tsc{WGM}
\tsc{QE}
\tsc{EP}
\tsc{PMS}
\tsc{BEC}
\tsc{DE}

\begin{document}
\begin{sloppypar}

\let\WriteBookmarks\relax
\def\floatpagepagefraction{1}
\def\textpagefraction{.001}
\shorttitle{Security Code Review with LLMs}
\shortauthors{J Yu et al.}
\title [mode = title]{\textcolor{black}{An Insight into Security Code Review with LLMs: Capabilities, Obstacles, and Influential Factors}}

\author[1]{Jiaxin Yu}
\ead{jiaxinyu@whu.edu.cn}
\credit{Conceptualization, Methodology, Investigation, Data curation, Formal analysis, Writing - Original draft preparation}

\author[1]{Peng Liang}
\cormark[1]
\ead{liangp@whu.edu.cn}
\credit{Conceptualization, Methodology, Investigation, Data curation, Supervision, Writing - Original draft preparation}
\address[1]{School of Computer Science, Wuhan University, China}

\author[1]{Yujia Fu}
\ead{yujia_fu@whu.edu.cn}
\credit{Investigation, Data curation, Writing - review and editing}

\author[2]{Amjed Tahir}
\ead{a.tahir@massey.ac.nz}
\credit{Conceptualization, Methodology, Writing - review and editing}
\address[2]{School of Mathematical and Computational Sciences, Massey University}

\author[3]{Mojtaba Shahin}
\ead{mojtaba.shahin@rmit.edu.au}
\credit{Conceptualization, Methodology, Writing - review and editing}
\address[3]{School of Computing Technologies, RMIT University, Australia}

\author[1]{Chong Wang}
\cormark[1]
\ead{cwang@whu.edu.cn}
\credit{Conceptualization, Methodology, Writing - review and editing}

\author[1]{Yangxiao Cai}
\ead{yangxiaocai@whu.edu.cn}
\credit{Investigation, Data curation}

\cortext[cor1]{Corresponding author.}


\begin{abstract}
Security code review is a time-consuming and labor-intensive process typically requiring integration with automated security defect detection tools. However, existing security analysis tools struggle with poor generalization, high false positive rates, and coarse detection granularity. Large Language Models (LLMs) have been considered promising candidates for addressing those challenges. In this study, we conducted an empirical study to explore the potential of LLMs in detecting security defects during code review. Specifically, we evaluated the performance of \black{seven} LLMs under five different prompts and compared them with state-of-the-art static analysis tools. We also performed linguistic and regression analyses for \black{the two top-performing LLMs} to identify quality problems in their responses and factors influencing their performance. 
Our findings show that: (1) In security code review, LLMs significantly outperform state-of-the-art static analysis tools, and the reasoning-optimized LLM performs better than general-purpose LLMs.
(2) DeepSeek-R1 achieves the highest performance, followed by GPT-4 \black{provided in the ChatGPT platform}. The optimal prompt for DeepSeek-R1 incorporates both the commit message and chain-of-thought (CoT) guidance, while for \black{GPT-4 via ChatGPT}, the prompt with a Common Weakness Enumeration (CWE) list works best.
(3) \black{GPT-4 via ChatGPT} frequently produces vague expressions and exhibits difficulties in accurately following instructions in the prompts, while DeepSeek-R1 more commonly generates inaccurate code details in its outputs.
(4) LLMs are more adept at identifying security defects in code files that have fewer tokens and security-relevant annotations.
\black{(5) Higher code complexity correlates with enhanced detection capabilities of DeepSeek-R1 for specific security defect types.}
\end{abstract}


\maketitle

\section{Introduction}
\label{sec:introduction}
Security defects are potential security risks or weaknesses in the system introduced during implementation~\citep{paul2021security}. Once attackers maliciously exploit a security defect, it could lead to serious consequences, including data breaches, financial losses, and service disruptions~\citep{telang2007empirical,cavusoglu2004effect}. The longer a security defect stays in the program, the higher its associated fixing and maintenance costs are~\citep{planning2002economic}. Thus, many organizations are shifting security analysis to earlier stages of software development, typically at code review time~\citep{gitlab2022url}. Code review is a human-intensive process in which the reviewer examines the code submitted by the developer to detect bugs, verify the implementation of the specification, ensure compliance with guidelines, and ensure quality. During this process, the reviewer raises issues in the code, discusses them with the developer, and provides recommendations. Adopting security analysis in code review can incorporate diverse viewpoints from both reviewers and developers, and effectively prevent security defects created by programmers working alone, as they naturally have a limited perspective of potential security risks~\citep{weir2017d}, from being merged into the source code repository. In this study, we focus on security analysis in the context of code review, referring to it as \textit{security code review}~\citep{edmundson2013empirical}. 

Security code review is increasingly integrated into the development pipelines by project teams~\citep{paul2021security}. Its resource-intensive nature, which requires significant human effort and time to review and revise the code, poses a notable challenge, particularly in popular large-scale open-source projects with numerous contributions~\citep{guo2024exploring}. Therefore, it has been desired to develop effective automated tools to assist code reviewers in security code review. Despite various static/dynamic/hybrid program analysis tools being proposed, they each encounter practical challenges, such as excessive imprecision~\citep{nong2021evaluating}, high false positive rates~\citep{johnson2013don, sui2020recall}, input range limitations~\citep{haller2013dowsing}, and weak scalability~\citep{nong2021evaluating}. 

Several data-driven approaches based on machine learning (ML)/deep learning (DL) have been developed for vulnerability detection and repair, which can somewhat handle the challenges encountered by traditional program analysis tools~\citep{chakraborty2021deep}. However, due to the lack of sizable and realistic training datasets~\citep{nong2022generating}, those approaches inevitably lacked robustness and tended to fall short when implemented in unfamiliar, real-world projects~\citep{nong2024chain}. \cite{nong2024vgx} found that even using the training dataset augmented by the latest vulnerability injection technique, the highest F1 score and the top-1 accuracy that state-of-the-art DL-based vulnerability detection and repair tools can achieve on the dataset constructed from real-world projects are still limited, at 20.01\% and 21.05\% respectively. Therefore, it remains essential to further explore techniques with better generalization and robustness to address vulnerabilities in real-world projects.

Recently, pre-trained LLMs have demonstrated promising performances across a broad spectrum of software engineering tasks, such as program repair~\citep{jiang2023impact}, test generation~\citep{kang2023large, schafer2023empirical}, and specification generation~\citep{xie2023impact}. Such models can understand and generate explanatory natural language responses, and are thus considered a potential candidate for security code review. We expect that LLM can act as a code reviewer, not only identifying security defects in code files, but also explaining the specific defect scenarios in detail to developers in a human-readable way. Applying LLMs to support security code review can help detect security defects earlier and improve productivity by helping code reviewers check their code faster and more efficiently. However, applying LLMs in the specific field of security code review remains largely unexplored.

Many studies have focused on leveraging LLMs for vulnerability detection. \cite{zhou2024largelanguage} evaluated the capabilities of GPT-3.5 and GPT-4 in vulnerability detection under few-shot learning prompts, while \cite{purba2023software} compared LLMs with static analysis tools in detecting software vulnerabilities. These studies mainly conducted a coarse-grained assessment of LLMs in binary classification tasks to judge whether there were security defects in the code. They did not require LLMs to provide more detailed information, such as the location and type of vulnerability. Researchers have also developed prompting strategies to enhance the effectiveness of LLMs in vulnerability detection, such as the DL-based prompting framework proposed by \cite{yang2024dlap} and the vulnerability-semantic guided prompting approach formulated by \cite{nong2024chain}. However, the datasets utilized in these studies are synthetic and CVE datasets, which do not fully represent real-world codebases. 

\textbf{Motivated by the limitations mentioned above}, we aim to bridge the knowledge gap by conducting an empirical study to comprehensively explore the potential of LLMs in fine-grained security code review on a practical dataset constructed from real-world code repositories. Specifically, in this study, LLMs were requested to provide detailed information on the identified security defect, including its line number, type, specific description, and suggested fix. The dataset we utilized consists of 534 code review files obtained from four open-source projects (namely, OpenStack Nova and Neutron, and Qt Base and Creator). We evaluated the performance of LLMs, analyzed existing quality problems, and examined factors influencing their performance to offer insights into the real-world applicability and limitations of LLM in security code review.
First, we identified five prompt templates to enhance LLMs (based on the prompting strategies formulated by \cite{zhang2024prompt}). \black{Then, using these prompt templates, we evaluated seven LLMs of diverse types, scales, and access methods on their ability to detect security defects identified by human reviewers, and compared their performance against that of static analysis tools.} After collecting responses generated by the two best-performing LLMs with their respective optimal prompts, we manually extracted quality problems present in these responses. We also leveraged these responses to construct two separate cumulative link models, aiming to explore the impact of 11 factors on the performance of each LLM. Our \textbf{findings} reveal that: 
(1) LLMs outperform state-of-the-art static analysis tools in security code review. Among LLMs, the reasoning-optimized model surpasses general-purpose models.
(2) DeepSeek-R1 is the best-performing LLM, followed by \black{GPT-4 provided in the ChatGPT platform}. For DeepSeek-R1, the prompt including the commit message and CoT instruction is the most effective, while for \black{GPT-4 (ChatGPT)}, the prompt with a CWE list works best.
(3) \black{GPT-4 via ChatGPT} struggles with vague expressions and poor instruction-following, while DeepSeek-R1 is more prone to producing incorrect code details in its responses.
(4) LLMs detect security defects more effectively in code files with fewer tokens and security-related annotations.
\black{(5) Higher code complexity is associated with improved detection performance of DeepSeek-R1 for security defects other than memory-related ones.}
In summary, this study makes the following \textbf{contributions}:
(1) We conducted a fine-grained evaluation of the capability of popular and representative LLMs in security code review. 
(2) We measured the impact of the non-determinism of LLMs on the consistency of security code review. 
(3) We identified quality problems and quantified hallucinations in the responses generated by the two top-performing LLMs, aiming to highlight the challenges faced by LLMs in security code review.
(4) We carried out the first analysis of the factors that might influence the performance of LLMs in security code review.

\textbf{Paper Organization.} Section~\ref{sec:related-work} surveys the related work. Section~\ref{sec:method} describes the methodology employed in this study. Section~\ref{sec:result} presents the results, followed by a discussion of the implications of our study results in Section~\ref{sec:implications}. Section~\ref{sec:threat} clarifies the threats to validity. Finally, Section~\ref{sec:conclusions} concludes this work and outlines future directions. \black{The replication package has been publicly released~\citep{replpack}.}

\section{Methodology}
\label{sec:method}

\subsection{Research Questions}
\begin{figure*}
    \centering
    \includegraphics[width=.99\linewidth]{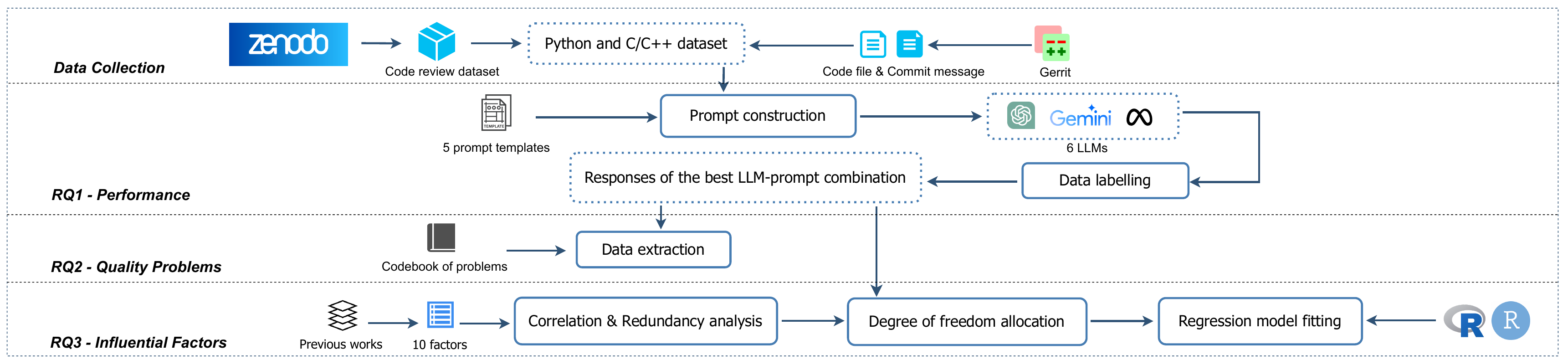}
    \caption{An overview of the research procedure for investigating the three RQs}  
    \label{fig:workflow}
\end{figure*}

This work aims to comprehensively assess the capability of LLMs to assist with security code review by automatically detecting security defects in the given code. We divided our study into three Research Questions (RQs), as shown below. The research procedure for each RQ is illustrated in Fig.~\ref{fig:workflow}.

\begin{tcolorbox}[arc=0mm,width=\columnwidth,
                  top=0mm,left=0mm,  right=0mm, bottom=0mm,
                  boxrule=.75pt]
\textbf{RQ1}: How effectively do LLMs detect security defects during code review? 
\end{tcolorbox}
\underline{RQ1.1}: How effectively do LLMs perform security code reviews under the basic prompt?

\underline{RQ1.2}: Can prompting with auxiliary information improve the performance of LLMs?

\underline{RQ1.3}: Can the Chain-of-Thought (CoT) prompting approach improve the performance of LLMs?

\underline{RQ1.4}: What is the impact of the non-determinism of LLMs on the consistency of security code review? 

\vspace*{0.3em}

\textbf{Motivation}. Many studies have focused on the performance of LLMs in detecting security defects, but they often suffer from low-quality datasets collected automatically and a narrow detection scope primarily confined to function-level code~\citep{zhou2025large}. To address these limitations, RQ1 utilizes a manually curated dataset to ensure data quality. It comprehensively evaluates the capability of LLMs to review a complete code file from real-world code reviews for security defects. Additionally, due to the inherent non-determinism of LLMs, the responses of LLMs may exhibit considerable divergence for the same prompt across different runs. Non-determinism is mainly caused by decoding configurations~\citep{song2024good}, with other hypotheses such as floating point, unreliable GPU calculations~\citep{ouyang2025empirical}, and its sparse MoE architecture failing to enforce per-sequence determinism~\citep{la2024language,puigcerver2023sparse}. Considering that the non-determinism of LLMs has not yet been studied when applied to security tasks, we conducted a quantitative analysis of the consistency of LLM responses to explore its impact on the task of security code review.

\begin{tcolorbox}[arc=0mm,width=\columnwidth,
                  top=0mm,left=0mm,  right=0mm, bottom=0mm,
                  boxrule=.75pt]
\textbf{RQ2}: \textcolor{black}{What quality problems exist in LLM-generated responses during security code review?}
\end{tcolorbox}
\underline{RQ2.1}: \textcolor{black}{What is the distribution of quality problems in these responses?}

\underline{RQ2.2}: \textcolor{black}{Which of these problems are hallucination-related, and what are their proportions? }

\textbf{Motivation}. \cite{kabir2023answers} noted that LLMs' responses may be verbose, inconsistent, and contain conceptual or logical errors, negatively impacting the quality of the answers. Among these problems, hallucination in particular has been extensively studied ~\citep{liu2024exploring, tonmoy2024comprehensive}. Hallucination refers to content generated by LLMs that appears plausible but lacks factual grounding~\citep{huang2025survey}. If we applied LLMs to real-world software development tasks, particularly during security code review, such hallucination may mislead practitioners, increase review costs, and even compromise system security. Previous studies have not yet thoroughly examined quality problems present in LLM-generated responses during security code review. RQ2 intends to conduct an in-depth linguistic analysis to investigate possible hallucination issues. First, we investigated the distribution of quality problems in LLMs' responses, such as \textit{Vague statement}, \textit{Incorrect concept}, and \textit{Incomplete answer}. Then we identified hallucination-related issues among these problems and quantified their proportion to bridge the knowledge gap and highlight potential areas for improvement.

\begin{tcolorbox}[arc=0mm,width=\columnwidth,
                  top=0mm,left=0mm,  right=0mm, bottom=0mm,
                  boxrule=.75pt]
\textbf{RQ3}: Which factors influence the performance of LLMs in security code review?
\end{tcolorbox}

\textbf{Motivation}. Previous studies have identified potential factors influencing the effectiveness of LLMs in detecting security defects~\citep{chen2023chatgpt, kabir2023answers, zhout2023devil}, primarily through case studies or observations. However, there is a lack of evidence substantiating such claims. RQ3 employs a regression analysis to validate whether these factors significantly impact the performance of LLMs. This RQ helps pinpoint areas of concern in applying LLMs to security code review and highlights possible improvement directions.
\begin{figure*}
    \centering
    \includegraphics[width=.99\linewidth]{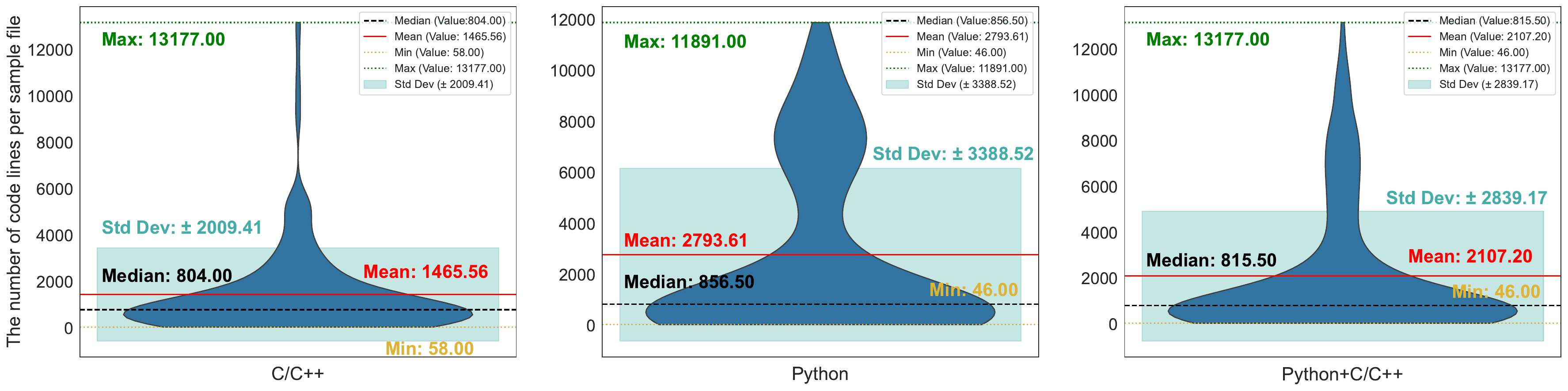}
    \caption{Distribution of LoC of the code files with security defects}   
    \label{fig:violin}
\end{figure*}

\begin{figure}
    \centering
    \includegraphics[width=.99\linewidth]{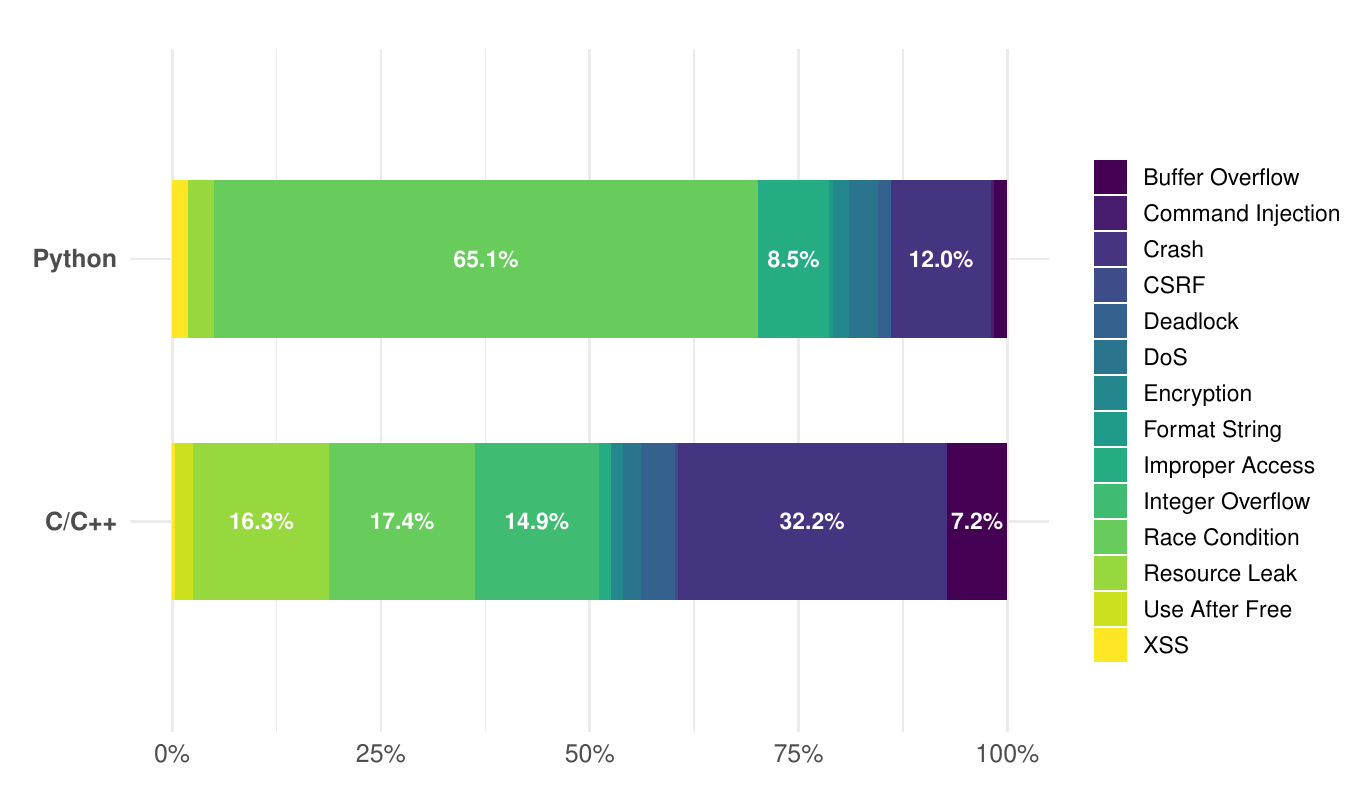}
    \caption{Distribution of security defect types on the Python and C/C++ datasets} 
    \label{fig:defect_type_distribution}
\end{figure}

\subsection{Data Collection}\label{sec:dataset}
We leveraged the code review dataset constructed by \cite{yu2023security} and examined the capability of LLMs in security code review. The dataset contains 614 review comments, each identifying a specific security defect across 15 predefined types from four projects: OpenStack Nova and Neutron (almost entirely written in Python), and Qt Base and Creator (mostly in C++ with some portions in C, Objective-C++, and Python). These projects are the most active within OpenStack and Qt~\citep{hirao2020code} and have been widely used in code review-related studies~\citep{han2022code, hamasaki2013does, spadini2018testing}. For each security defect, we could not include the code from the entire patchset in our prompt, which is always too long and exceeds the input token limits of LLMs. Instead, we included only the code from a single file in which a reviewer added a review comments to point out the security defect. The 614 security-related review comments were reorganized as follows:

First, we filtered out comments published at the patchset level, as such data cannot be located in a specific file to construct a prompt. Then, for each code file to which these comments correspond, we consolidated all security-related comments in the file and the corresponding security defect records into a single data item. Next, we utilized the Gerrit API to retrieve the content of each code file and the commit messages of the patchsets to which they belong. During this procedure, if we cannot obtain the content of the code files through the API, we ignore those files. Lastly, since 97\% of the files in this dataset are Python and C/C++ files, we decided to remove any other files (all are configuration and scripting files) as those only formed $\approx 3\%$ of the files and thus had a negligible impact on our results.   

Our final dataset consists of 534 code files (258 Python and 276 C/C++ files), each including source code, the commit message of the associated patchset, and detailed information of the \textbf{target security defect} as identified by the reviewers. We analyzed the distribution of security defect types in Python and C/C++ code files (see Fig.~\ref{fig:defect_type_distribution}). In Python code files, the Python language employs automatic memory management and garbage collection. Therefore, \textit{Memory Leak} and \textit{Buffer Overflow} are relatively rare, while \textit{Race Condition} is the most common type, accounting for 65.1\%. In contrast, C/C++ requires programmers to manage memory and pointers manually, thus the C/C++ dataset has a more balanced distribution of security defects. In C/C++ code files, \textit{Crash} is the most prevalent (32\%), followed by \textit{Race Condition} (17.4\%), \textit{Resource Leak} (16.3\%) and \textit{Integer Overflow} (14.9\%). We also counted the lines of code (LoC) of the collected code files in our dataset. Fig.~\ref{fig:violin} shows the distribution of LoC for different languages (Python and C/C++, respectively) and the distribution of LoC for the entire dataset. We found that the median LoC is 815 and the average reaches 2,107, which substantially exceeds the scale of function-level datasets like Big-Vul (avg. 30 LoC) and SVEN (avg. 168 LoC) adopted in related studies~\citep{yin2024multitask,steenhoek2024err}.

\subsection{Research Procedure of RQ1}
\begin{table*}[ht]
\caption{Versions and hyper-parameters of 6 LLMs studied}
\label{tbl:llms}
\scriptsize
\renewcommand{\arraystretch}{1.1}
\begin{tabular}{p{2.5cm}|p{2.2cm}|c|c|c|p{5cm}}
 \hline
        \multirow{2}{*}{\textbf{Model}}      &\multirow{2}{*}{\textbf{Version}}  &\multirow{2}{*}{\textbf{Context Window}}     &\multicolumn{3}{c}{\textbf{Parameter}}    \\\cline{4-6}
        ~                   &                  &          & temperature    & top\_p    &\multicolumn{1}{c}{\textbf{Other parameters}}\\ \hline
        GPT-4 (ChatGPT)               &   May 2024 version       &-                   & \multicolumn{3}{c}{-} \\ \hline
        GPT-4 (API)              &   gpt-4-0613       &8k                   & 1.0 &1.0            &max\_tokens: 2048; frequency\_penalty: 0.0; \\ \hline
        GPT-4 Turbo         &   gpt-4-1106-preview     &128k                    & 1.0 &1.0            &max\_tokens: 4096; frequency\_penalty: 0.0; \\ \hline
        Gemini Pro          &   gemini-1.0-pro         & 32k                  & 0.9 &1.0             &maxOutputTokens: 2048; \\ \hline
        Llama 2 7B          &   llama-2-7b-chat        & 4k                  & 1.0 &0.9              &max\_seq\_len: 4096; \\ \hline
        Llama 2 70B         &   llama-2-70b-chat       & 4k                 & 1.0 &0.9              &max\_seq\_len: 4096; \\ \hline
        DeepSeek-R1         &   deepseek-r1-250528     & 128k                 & 1.0 &0.7              &max\_tokens: 4096; frequency\_penalty:0.0;\\ \hline
        
\end{tabular}
\end{table*}
\subsubsection{LLM Selection}

\black{To fully evaluate the performance of LLMs in security defect detection, we selected a range of representative LLMs from both proprietary and open-source families. Specifically, this study encompasses three proprietary LLMs--GPT-4,  GPT-4 Turbo and Gemini Pro--as well as two open-source LLMs from the Llama 2 series, namely Llama 2 7B and Llama 2 70B. These models were the most widely adopted LLMs when we started our experiments in December 2023. Given the rapid development of LLMs and the emergence of reasoning-optimized models, we additionally employed DeepSeek-R1 and accessed its updated version released on May 28, 2025, to ensure the comprehensiveness of our evaluation. }

\black{Regarding deployment, GPT-4, GPT-4 Turbo, Gemini Pro and DeepSeek-R1 were accessed via their respective APIs. For Llama 2 7B and Llama 2 70B, we deployed chat versions of them on a server equipped with eight NVIDIA GeForce RTX 4090 GPUs, each with 24 GB of VRAM. To better reflect the practical capabilities of off-the-shelf LLMs in security defect detection, we used the default hyperparameter settings throughout our experiments. Although some studies mitigate the non-determinism of LLMs by setting the temperature parameter to zero~\citep{atil2024llm, blackwell2024towards}, we refrained from doing so, as this may adversely affect the performance of the LLMs~\citep{song2024good} and may not guarantee determinism~\citep{ouyang2025empirical, astekin2024exploratory}. We retained the default temperature settings and quantified the non-determinism of LLMs as one aspect of performance evaluation in RQ1.4, rather than artificially suppressing such non-determinism via hyperparameter tuning. The versions, context windows, and parameter settings of LLMs used in our study are summarized in Table \ref{tbl:llms}.}

Notably, \black{GPT series models can be accessed through both the API and the ChatGPT platform, and the performance of the same model may vary depending on the access method. In the preliminary exploration stage, we compared the performance of GPT-4 accessed through these two ways in security code review and found that GPT-4 provided by the ChatGPT platform outperformed the one accessed through the API. Therefore, we also included GPT-4 provided by the ChatGPT platform (May 2024 version) in our performance comparison. During the experiments, to obtain outputs from the ChatGPT platform, a new chat session was initiated for each sample without any prior conversations, chat history and personalized settings, minimizing the potential impact of account-level state on model performance.} As the default parameters of \black{GPT-4 via ChatGPT} were not publicly disclosed, the corresponding entries in Table \ref{tbl:llms} are indicated with `-'.

\subsubsection{Prompt Design}
\label{sec: prompt_design}

\begin{figure*}[htbp]
        \centering
        \includegraphics[width=0.95\linewidth]{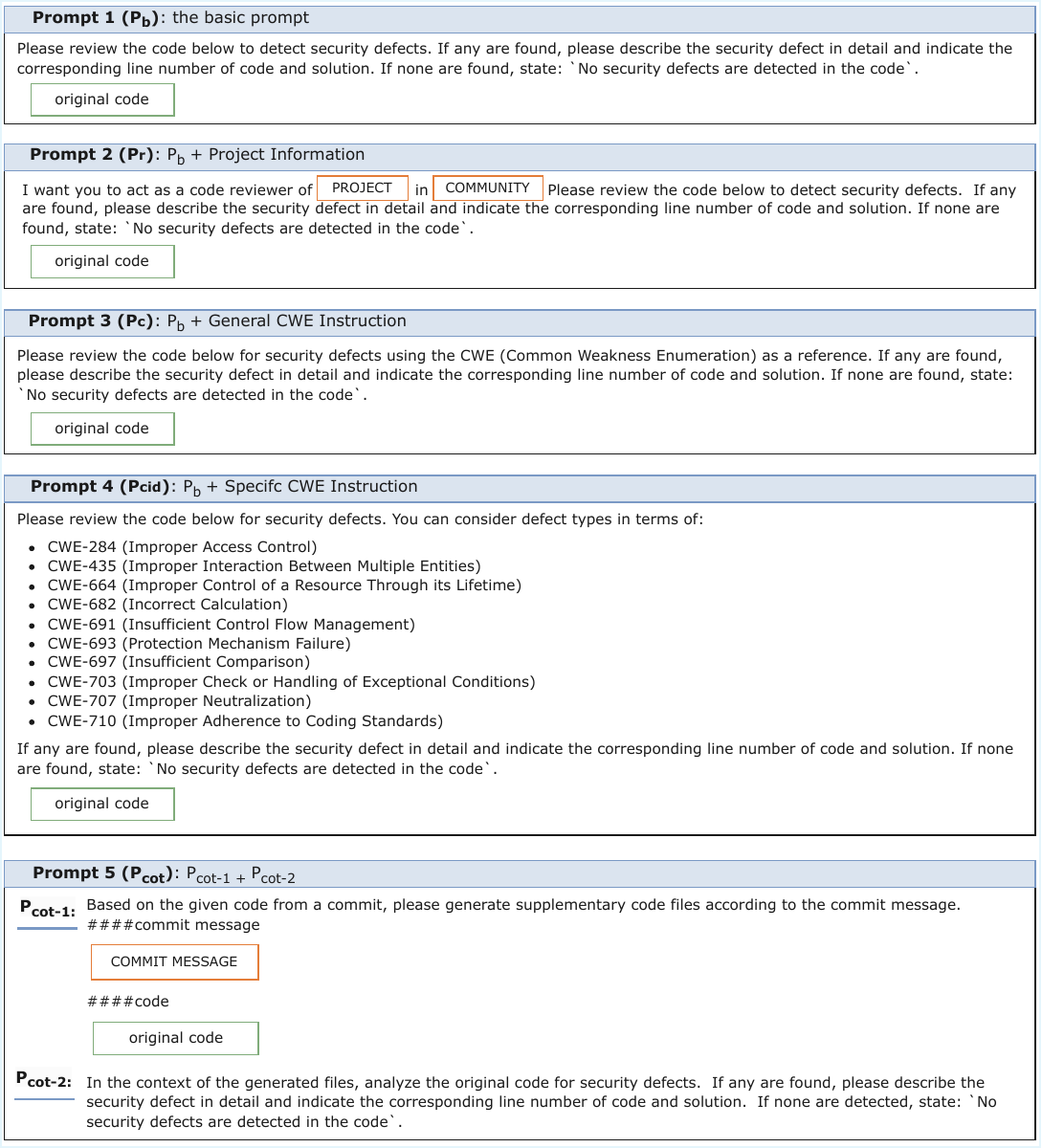}
        \caption{Construction templates for the five prompts}
        \label{fig:prompt1}  
        \vspace{0.5em} 
        \centering
        \includegraphics[width=0.95\linewidth]{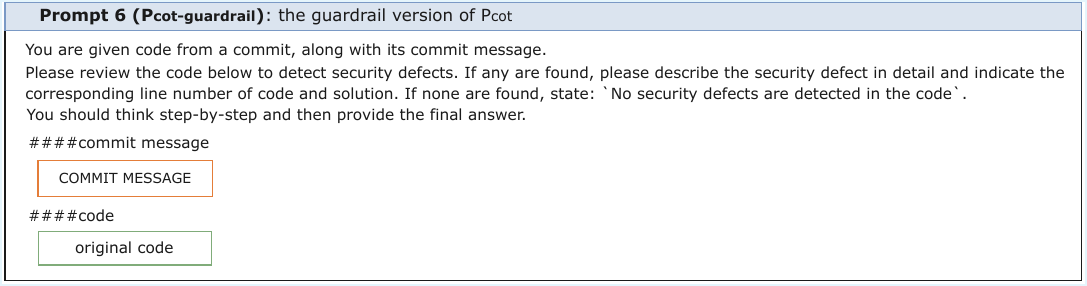}
        \caption{Construction template for the guardrail version of \(\textbf{P}_{cot}\)}
        \label{fig:prompt6}  
\end{figure*}

Prompt engineering is a process of designing prompts to optimize model performance on downstream tasks~\citep{liu2023pre}. To design prompts, we followed the best practices published by OpenAI~\citep{gpt2023bestpractices} and the three prompt design strategies in \cite{zhang2024prompt}: basic prompt, enhanced prompt with auxiliary information, and CoT prompt. During our preliminary exploration stage, we tried prompts with various auxiliary information and different CoT intermediate reasoning steps. To fully harness the capabilities of LLMs in detecting security defects, we finally selected the five best-performing prompts. Fig.~\ref{fig:prompt1} illustrates the construction strategies for the five prompts.

\begin{itemize}[leftmargin=*, label=\textendash, itemsep=-0.7ex]
    \item \textbf{Prompt 1 (\(\textbf{P}_b\)): the basic prompt}. We instruct the LLM to review the provided code for security defects and output their descriptions, code line numbers, and suggested fixes. If no defects are found, the LLM should output a fixed sentence, `\textit{No security defects are detected in the code.}'. It is worth noting that we instruct the model to generate a solution for each identified security defect, since an incorrectly detected security defect usually cannot lead to a logically consistent fix in the source code. This process naturally drives the model to evaluate the validity of its detection and acts as an implicit self-reflection mechanism.
    \item \textbf{Prompt 2 (\(\textbf{P}_r\)): \(\textbf{P}_b\) + Project Information}. \(\textbf{P}_r\) is designed based on \(\textbf{P}_b\) but includes the name of the source project as auxiliary information. The LLM is asked to act as a code reviewer for the project using the provided code in the prompt.
    \item \textbf{Prompt 3 (\(\textbf{P}_c\)): \(\textbf{P}_b\) + General CWE Instruction}. \(\textbf{P}_c\) directly instructs the LLM to use CWE as a reference for identifying security defects without providing specific CWE details.
    \item \textbf{Prompt4 (\(\textbf{P}_{cid}\)): \(\textbf{P}_b\) + Specific CWE Instruction}. \(\textbf{P}_{cid}\) provides a specific list of the first level CWEs in VIEW-1000 (Research Concepts)\footnote{VIEW-1000 is a weaknesses classification framework designed for academic research, widely adopted in prior work (e.g., ~\citep{pan2023fine}). It organizes CWEs into a hierarchical structure based on parent-child relationships among CWE entries. The first level of CWEs in VIEW-1000 encompasses a broad range of commonly encountered security defects, thus applied in Prompt 4 (\(\textbf{P}_{cid}\))}. We acknowledge that providing a list of CWEs may limit the detection scope and miss security defects that do not fall into the provided CWE categories. However, including all CWEs is infeasible due to the limited context window of the LLM and the potential loss of focus.
    \item \textbf{Prompt 5 (\(\textbf{P}_{cot}\)): \(\textbf{P}_{cot-1}\)+\(\textbf{P}_{cot-2}\)}. Since we only included the code of a single file in the prompt instead of an entire patchset, we designed \(\textbf{P}_{cot}\) to analyze the impact of missing contextual information on the performance of LLMs. The task of security code review is split into two steps. In \(\textbf{P}_{cot-1}\), given that we only provide a single code file from a complete project in the prompt, we include the corresponding commit message of the file and instruct the LLM to generate code context according to the commit message. Then \(\textbf{P}_{cot-2}\) guides the LLM to review the provided code for security defects with reference to the generated code context.
\end{itemize}
However, in the two-stage \(\textbf{P}_{cot}\), the performance of LLMs can be influenced by three factors: the commit message, CoT guidance, and code context generated by the LLM. Among these, the fabricated code context may contradict the actual code, leading to incorrect judgments. To isolate the impact of the LLM-generated context, we proposed a guardrail version of the \(\textbf{P}_{cot}\) prompt and conducted experiments exclusively on DeepSeek-R1 under this additional prompt. 
\begin{itemize}[leftmargin=*, label=\textendash, itemsep=-0.5ex]
    \item \textbf{Prompt 6 (\(\textbf{P}_{cot-guardrail}\)):} This prompt retains the commit message and CoT guidance but does not use code context generated by the LLM, allowing us to evaluate the effect of fabricated contexts. The detailed template of \(\textbf{P}_{cot-guardrail}\) is shown in Fig~\ref{fig:prompt6}.
\end{itemize}

\subsubsection{Evaluation Metrics}
\label{sec:metrics}

During the experiments, responses of LLMs were provided in the form of natural language text, making it necessary to manually inspect these responses to ensure the accuracy of the performance measurements. We considered that LLM-generated responses could be evaluated in a manner similar to the approach adopted by \cite{mahajan2020recommending} for measuring responses in Stack Overflow. We adapted the evaluation metrics formulated in their work~\citep{mahajan2020recommending} as follows: 
First, we excluded the `Unavailable' category defined by \cite{mahajan2020recommending} because LLMs do not have a case of no response. Next, the definitions of each category were adjusted to match the task of security code review. Specifically, each response was categorized into one of the four rating categories, i.e., `Instrumental', `Helpful', `Misleading' and `Uncertain'. To ensure clearer differentiation, for the `Helpful' and `Instrumental' categories, we specified their definitions based on their performance across three key attributes, i.e., defect type, location, and description. The detailed definitions and corresponding examples of these four rating categories are as follows:

\black{\textbf{Instrumental (\textsf{I})}: The type of the security defect identified in the response is exactly the same as the target security defect, the location is strictly within the code line range of the target security defect, and the defect description can provide effective information for understanding the scenario.}

\begin{tcolorbox}[colback=blue!5!white, colframe=blue!75!black, fontupper=\small, title=Example 1 -  Instrumental]
\textbf{Reviewer:} (Line 76) I think this is race-y. Let's say we have a three coroutines scheduled to call function. the first may yield at L74, the second enters before the threshold and sleeps at L70 while setting is\_deferred, then the first finishes and unsets at L76, subsequently allowing the third to enter before the threshold as well. Then both, second and third coroutines call function in very short time difference.
\newline\newline
\textbf{LLM:} Race condition in `Throttler.call()`(Lines 58-77):
\newline - The is\_deferred flag is used to determine if a function should be deferred. However, multiple threads can check the is\_deferred flag almost simultaneously and hence more than one thread might end up executing the contained code.
\newline\newline
\textit{\textbf{Explanation:} LLM indicated the race condition defect identified by the reviewer, with the accurate code line range and detailed description, thus categorized as \textbf{Instrumental}.}
\end{tcolorbox}
\vspace*{1mm}

\black{\textbf{Helpful (\textsf{H})}: The type or location of the identified security defect differ to some extent from the target security defect, or the description may be too broad or partially inaccurate. However, the response still provides useful evidence pointing toward the existence of the target defect.}
\vspace*{1mm}
\begin{tcolorbox}[colback=blue!5!white, colframe=blue!75!black, fontupper=\small,title=Example 2 - Helpful]
\textbf{Reviewer:} (Line 141) If connect\_info ever includes a secret of a type other than token (e.g. password) that mask\_password looks for, you'd be replacing that with the masked token instead of masking the password or whatever it is. But I guess that's unlikely.
\newline\newline
\textbf{LLM:}Potential Information Disclosure(Line 141):
\newline - Logging connection information can potentially expose sensitive details. Even though `mask\_password` is being used, it's essential to validate that no other sensitive information is leaked.
\newline\newline
\textit{\textbf{Explanation:} The reviewer noted that the mask\_password function might incorrectly replace different types of sensitive information. While the LLM correctly identified the relevant, it only provided general emphasis on checks for potential sensitive information exposure, failing to recognize this specific risk scenario. So this instance was categorized as \textbf{Helpful} rather than \textbf{Instrumental}.}
\end{tcolorbox}
\vspace*{1mm}
\black{\textbf{Misleading (\textsf{M})}: The response does not provide any helpful information, e.g., claiming no defects were found or failing to perform the detection task.}
\vspace*{1mm}
\begin{tcolorbox}[colback=blue!5!white, colframe=blue!75!black, fontupper=\small, title=Example 3 -  Misleading]
\textbf{Reviewer:} (Line 329) This does implicit format detection on the imported disk, which is generally considered a well-known security hole. In this specific case a malicious user could create a `raw' image which actually contained a malicious qcow2 header specifying a backing file of the the compute host's root disk. This would result in copying the compute host's root disk to the specified rbd volume and exposing it to the user.
\newline\newline
\textbf{LLM:} Based on the provided code snippet, no security defects are detected. For a more comprehensive analysis, a full code audit in the context of the entire application and its environment would be recommended.
\newline\newline
\textit{\textbf{Explanation:} The reviewer pointed out a sensitive data exposure at line 329, whereas the LLM did not detect any security defects. Therefore, it was classified as \textbf{Misleading}.}
\end{tcolorbox}
\vspace*{1mm}
\black{\textbf{Uncertain (\textsf{U})}: The response does not mention any defects identified by the reviewer but instead points out the existence of other security defects. Due to the lack of expertise of the researchers about the code and its context, the actual existence of these security defects cannot be confirmed.}

\begin{tcolorbox}[colback=blue!5!white, colframe=blue!75!black, fontupper=\small,title=Example 4 -  Uncertain]
\textbf{Reviewer:} (Line 469) So not even timeout cancels the future? Will this cause that threads will hang around forever leaking OS resources? This feels dangerous.
\newline\newline
\textbf{LLM:} **Potential Insecure Direct Object References (Lines 89-563):**
\newline - The `RequestContext` class and its methods provide direct access to project and user identifiers, and various methods in this file allow direct manipulation of database and message queue connections. If access controls are not properly enforced or if input validation is weak, this could lead to unauthorized access or manipulation of data.
\newline\newline
\textit{\textbf{Explanation:} The reviewer identified that the thread handling may lead to OS resource leaks. However, LLM did not mention any about resource leak, instead highlighting a potential improper access issue. Since access mechanisms might be uniformly managed by a middleware, and considering the unavailability of detailed implementation information, we cannot determine whether the issue pointed out by LLM exists. Thus this instance was categorized as \textbf{Uncertain}.}
\end{tcolorbox}
\vspace*{1mm}
The corresponding evaluation metrics are I-Score (\(\frac{I}{I+H+U+M} \times 100\%\)), IH-Score (\(\frac{I+H}{I+H+U+M} \times 100\%\)), and  M-Score (\(\frac{M}{I+H+U+M} \times 100\%\)). Higher I-Score and IH-Score indicate a stronger ability of the model to detect security defects identified by reviewers. At the same time, a lower M-Score signifies fewer misleading results produced during security defect detection. Given that LLMs exhibit considerable non-determinism, we repeated the experiment for each LLM-prompt combination three times and calculated the average of the three performance scores as the final result, so as to alleviate this threat and ensure a reliable evaluation of LLMs' capabilities in security defect detection.

It is important to note that the cost and latency of LLMs in security code review are significant concerns for practitioners. However, in this study, these two points were not specifically discussed. Due to the various invocation methods of the LLMs involved (including calls through the ChatGPT platform, local deployment, and API calls), it is difficult to provide a consistent cost estimate across different LLMs. Moreover, the latency of the models is significantly affected by their deployment environments, such as computational resources, inference engines, and parallelization strategies, making the latency of different LLMs in our experiments incomparable.

\subsubsection{Baselines}
We tried different tools and learning techniques for detecting security defects to select appropriate baselines for our work.To compare with LLMs, we focused on techniques providing fine-grained detection, including classification, localization, and description of defects. This led us to exclude techniques with insufficient detection granularity, such as ProRLearn~\citep{ren2024prorlearn}, GRACE~\citep{lu2024grace} and LineVul~\citep{fu2022linevul}. Given that we evaluated the performance of LLMs on the C/C++ and Python datasets separately, for language-specific static analysis tools, we selected CppCheck~\citep{cppcheck} for C/C++ and Bandit~\citep{bandit} for Python.
Furthermore, we also used CodeQL~\citep{codeql} and SonarQube~\citep{sonarqube} to analyze Python files. These two tools can directly analyze code files written in dynamically-typed languages like Python but require code files in statically-typed languages like C/C++ to be compiled first. Therefore, we could not apply them to our C/C++ dataset, which is comprised of uncompiled and isolated C/C++ code files.
Additionally, since Semgrep~\citep{semgrep} supports multiple programming languages and can analyze individual code files, we adopted it for both C/C++ and Python datasets. To obtain comprehensive analysis results, CodeQL adopted the \texttt{Python-security-and-quality.qls} test suite, covering a wide range of issues from basic code structure and naming conventions to advanced security and performance vulnerabilities. Other tools utilized their default rule sets for analysis.

\subsubsection{Data Labelling}
Prompts were constructed using five templates and then fed into seven LLMs to collect their responses. Since the responses of LLMs were provided in the form of natural language text, to rate the responses accurately, we manually inspected the content of LLM-generated responses, their corresponding source code, and security defects identified by reviewers to categorize them into the four categories defined in Section~\ref{sec:metrics}.

To mitigate bias, pilot data labeling was conducted, in which we randomly selected 100 out of 534 code files from our dataset, constructed five prompts for each file and fed these prompts into LLMs to collect responses. Using the definitions of the four categories as labeling criteria, the first and third authors labeled the collected responses separately. Discrepancies in labeling results were discussed with the second author to reach a consensus. The inter-rater reliability was calculated using Cohen's Kappa coefficient~\citep{cohen1960coefficient}, yielding a value of 0.89. Then, the first and third authors labeled all the rest of the responses and discussed any ambiguous cases with the second author to finalize the categorization. For the additional experiments on the combination of \(\textbf{P}_{cot-guardrail}\) and DeepSeek-R1, we applied the same labeling approach. Based on the labeling results, we calculated the performance scores of each LLM-prompt combination in the Python and C/C++ datasets. 
We found that the best-performing LLM is DeepSeek-R1, followed by \black{GPT-4 (ChatGPT)}. On the complete dataset (including both Python and C/C++ files), the optimal prompts for DeepSeek-R1 and \black{GPT-4 (ChatGPT)} are \(\textbf{P}_{cot-guardrail}\) and \(\textbf{P}_{cid}\), respectively, as detailed in Section~\ref{sec: rq1_result}. \black{Therefore, for further analysis of RQ2 and RQ3, we selected the responses generated by \textit{DeepSeek-R1+\(\textbf{P}_{cot-guardrail}\)} and \textit{GPT-4 (ChatGPT)+\(\textbf{P}_{cid}\)} on the complete dataset across three repetitive experiments. Since some code files exceeded the token limit, we ultimately obtained 1,509 responses for DeepSeek-R1 and 1,263 responses for GPT-4 (ChatGPT), with each experiment contributing 503 and 421 responses, respectively.}

\subsubsection{Consistency Calculation}
To measure the non-determinism of LLMs, we adopted the metrics in \cite{chang2024beyond} and utilized entropy to measure the consistency of responses generated by LLMs across three repetitive experiments. Specifically, an LLM generated three responses for a given code file under a specific prompt template. We then calculated the entropy for the four response categories in the three responses. We determined the overall consistency by averaging the entropy across all code files using the following formula:
\(\frac{1}{R} \sum_{r \in R} \text{entropy}(p^{r}_{1}, \dots, p^{r}_{4})\), where R is the response count in each round of the experiment. A higher mean entropy suggests the LLM has higher inconsistency in repetitions of experiments. To further quantify the uncertainty of the entropy estimates and ensure the reliability of our findings, bootstrap confidence intervals were calculated for each LLM–prompt combination. Following the guidance of ~\cite{efron1994introduction}, we resampled the original dataset with replacement to match its original size and calculated the mean entropy for each resample. As suggested by ~\cite{davison1997bootstrap}, if confidence levels 0.95 and 0.99 are to be used, then it is advisable to have [the number of resamples]=999 or more, if practically feasible. Therefore, consistent with the settings adopted in ~\cite{ahmed2025can}, we performed 1,000 bootstrap replications to obtain the 95\% confidence intervals for the entropy estimates.

\subsection{Research Procedure of RQ2}
We manually inspected the responses generated by \black{\textit{DeepSeek-R1+\(\textbf{P}_{cot-guardrail}\)} and \textit{\black{GPT-4 (ChatGPT)}+\(\textbf{P}_{cid}\)}} across three repetitive experiments to analyze the quality problems present in these responses. We employed the open coding and constant comparison~\citep{glaser1965constant} with predefined categories by the MAXQDA tool for qualitatively analyzing the responses.

Based on the problem categories formulated by \cite{kabir2023answers} and the problems of LLM-generated text in various domains identified by previous studies~\citep{borji2023categorical,mitrovic2023chatgpt,kalla2023advantages}, two of the authors firstly inspected a few responses independently and recorded their observations. Then they collaboratively reviewed the inspected responses and discussed to establish an initial codebook, resulting in four themes: \textbf{Correctness}, \textbf{Understandability}, \textbf{Conciseness}, and \textbf{Compliance}, each with a series of problem types. In particular, we excluded the \textit{Consistency} theme defined in \cite{kabir2023answers} from our study. The reason is that the consistency between the LLM's response and the reviewer's assessment (i.e., the ground truth) has been investigated in RQ1. Following the codebook, we conducted a pilot data analysis across two iterations to improve the agreement between the two authors. In each round, ten responses were randomly selected for the two authors to analyze their quality problems. The results were then compared and discussed, with partial adjustments made to the codebook. Since a single response could contain multiple problems, two authors completed the remaining data analysis task where the labels are not mutually exclusive. Therefore, we calculated the agreement level using Krippendorff’s alpha~\citep{krippendorff2004measuring} with Jaccard distance, which was improved from 0.40 in the first round to 0.80 in the second, indicating a high level of agreement. Then the two authors completed the remaining data analysis and verified by the second author. Throughout the entire procedure, we employed a negotiated agreement approach~\citep{campbell2013coding}: any conflicts were consulted and addressed by three authors, ensuring the reliability of data analysis. For each quality problem type, we computed its average frequency across three repetitive experiments of two LLM+prompt combinatons, i.e.,  \textit{DeepSeek-R1+\(\textbf{P}_{cot-guardrail}\)} and \textit{\black{GPT-4 (ChatGPT)}+\(\textbf{P}_{cid}\)}, as the final occurrence rate of this problem in the responses of each combination.

We consider the finalized codebook to cover all types of quality problems present in the LLM-generated responses within our dataset. To further analyze the proportion of hallucinations in LLMs' responses, we adopted the granular taxonomy proposed by~\cite{huang2025survey}, which categorizes hallucinations into two main types: Factual and Faithfulness. \textbf{Factual Hallucination} refers to outputs that contradict real-world facts or lack factual grounding, including \textit{Factual Contradiction} and \textit{Factual Fabrication}. \textbf{Faithfulness hallucination}, on the other hand, refers to deviations of the output from the instruction, current context, and reasoning process, including \textit{Instruction Inconsistency}, \textit{Context Inconsistency}, and \textit{Logical Inconsistency}. We mapped the quality problems defined in our codebook to these hallucination subtypes to calculate their distribution across the dataset. Detailed mapping and results are provided in Section ~\ref{sec: rq2_result}.

\subsection{Research Procedure of RQ3}
To answer RQ3, we adopted the approach formulated by Harrell~\citep{harrell2001regression} to construct a regression model. As mentioned in Section~\ref{sec:metrics}, the response variables in our dataset are rated as \textsf{I}, \textsf{H}, \textsf{U}, and \textsf{M}. In terms of the helpfulness of the response, by definition, \textsf{I} is more important than \textsf{H}, and so forth. Accounting for the ordinal nature of the response variable, we ultimately aggregated the data from all three experiments to fit a \black{cumulative link model~\citep{powers2008statistical}, which supports ordinal and categorical response variables.} The specific construction process is described in the following subsections.

\begin{table*}[ht]
\caption{Factors that may influence the performance of LLM in security code review}
\label{tbl:factors}
\scriptsize
\setlength{\tabcolsep}{4pt}
\begin{threeparttable}
  \begin{tabular}{ |p{3cm}|p{3cm}|p{5cm}|p{5cm}| } 
    \hline
    \textbf{Factor}             &\textbf{Source*}                                                &\textbf{Definition}                            &\textbf{Rationale}              \\\hline
    Token\newline({TK})           & Adjusted from ~\cite{paul2021security, mcintosh2016empirical} & Number of tokens of the code file. Computed using the TikToken~\citep{tiktoken} tokenizer.             & Larger code files may be more difficult for LLMs to analyze. \\\hline

    Position\newline({PT})                      &Collected from ~\cite{Sovrano_2025} & Position of the security defect within the code file. Determined by calculating the proportion of tokens preceding the defect relative to the total token count, using the TikToken~\citep{tiktoken} tokenizer.             & LLMs tend to significantly underperform when detecting security defects located toward the end of files~\citep{Sovrano_2025, rafi2024order}. \\\hline
    
    FileType\newline({FT})                 & Created                                                       & Type of the code file, i.e., \textit{Source} or \textit{Auxiliary}. \textit{Source} indicates files that directly contain the logic of the application (i.e., \texttt{.cpp}, \texttt{.c} and \texttt{.py}) while \textit{Auxiliary} files are utilized for declarations (i.e., \texttt{.h} and \texttt{.hpp}).                       & The understanding capability of LLMs and the scenarios of security defects may vary across different types of files.                        \\\hline
    
    SecurityDefectType\newline({SDT})         &  Transformed from ~\cite{chen2023chatgpt}                       & Type of the security defect that the reviewers identified in this code file. \black{Due to the scarcity of samples for certain security defect types, we consolidate the original 15 types into five broader categories, abbreviated as \textit{Thread}, \textit{Memory}, \textit{Resource}, \textit{Crash}, and \textit{Permission} (representing -related defects).} & ChatGPT is proven to display distinct detection performances across various vulnerabilities~\cite{chen2023chatgpt, zhout2023devil}. Thus, we assume that LLMs may be more adept at detecting certain specific types of security defects in code.                           \\\hline                                                   
    Commit\newline({CT})                       & Collected from ~\cite{paul2021security}                                      & Number of historical commits for the code file in its code change.  & Excessive history commits may lead developers to fatigue or overconfidence, resulting in their ignorance of security considerations in simpler scenarios.                        \\\hline
    Complexity\newline({CPT})                  & Collected from ~\cite{paul2021security, mcintosh2016empirical}               & McCabe’s Cyclomatic Complexity~\cite{mccabe1976complexity}. Computed using the Lizard~\cite{lizard} analyzer. & Code comprehensibility may become more challenging for LLMs as its cyclomatic complexity increases~\cite{paul2021security}.                        \\\hline
    Community\newline({CMT})                   & Created                                                       & Name of the community source of the code file, i.e., \textit{OpenStack} or \textit{Qt}.  &OpenStack and Qt differ significantly in terms of their structure, technology stack, project function, and main components. The LLMs' performance may differ on data derived from these two communities.\\\hline
    AuthorExperience\newline({AE})           & Collected from ~\cite{paul2021security}                                       & Number of closed code commits the author has submitted prior to the commit this code file corresponding to.   & Experienced authors may write more rigorous and standardized code, while the security vulnerabilities in the code are also relatively more subtle and difficult to detect.                        \\\hline
    AnnotationRatio\newline({AR})       & Created                                       & The ratio of manual annotations in the code file. If the token number of all annotations in the file \textbf{f} is \(\textbf{a}_{t}\) and the total token number of this file is \(\textbf{f}_{t}\), then \(\textbf{AR} = \textbf{a}_{t}/\textbf{f}_{t}\)    &  The capability of LLMs to comprehend manual annotations and code may vary.                        \\\hline
    AnnotationHasCode\newline({AHC})         & Transformed from ~\cite{chen2023chatgpt}                        & Whether the annotations in the code file include code.   & Code in annotations was abandoned by developers but often interpreted as actual code by LLMs~\cite{chen2023chatgpt}, leading to incorrect security defect detection.\\\hline
    AnnotationisSecurityRelated\newline({ASR}) & Created  & The relevance of annotations to the security defects identified by reviewers in this code file, i.e., 1 (annotations mentioned target security defects identified by reviewers), 0 (Annotations were unrelated to security) and -1 (Annotations mentioned non-target security defects).    & Keywords in the description related to target security defect in annotations can act as cues for LLMs, while the descriptions of non-target security defects often involve the historical fixing or security mechanisms, which may cause interference to LLMs.                        \\\hline
\end{tabular}
\begin{tablenotes}
    \item[\textbf{*}] The sources of factors correspond to the descriptions in \textbf{bold} in Section~\ref{sec:ac}.
\end{tablenotes}
\vspace*{-1em}
\end{threeparttable}
\end{table*}
\subsubsection{Attributes Construction} \label{sec:ac}

To comprehensively collect factors influencing the detection of security defects by LLMs, excluding the prompt and the model itself, we undertook the following steps: First, we \textbf{collected} the attributes from previous works~\citep{paul2021security, mcintosh2016empirical, thongtanunam2017review} that were utilized in analyzing relationships related to code review and applicable to the code file data in our work.
Given that LLMs understand and generate texts token by token, we \textbf{adjusted} the factor adopted from \cite{mcintosh2016empirical} --- `Size', to `Token' (see the first row of Table~\ref{tbl:factors}), as larger code files typically contain more tokens. Next, we extracted characteristics of data that were explicitly indicated to affect or correlate with the capabilities of LLMs in \cite{chen2023chatgpt, Sovrano_2025}, \textbf{transforming} them into factors for our experiment. Lastly, after thoroughly observing and analyzing the distribution of responses, we further \textbf{created} and adopted several potentially influential factors based on our observations. In the end, we obtained a list of 11 factors in Table~\ref{tbl:factors}, along with the source, definition, and rationale for each factor. 

Since acquiring a factor like `AuthorExperience' entails querying the code review history of projects, we utilized Gerrit API to retrieve detailed information for all patchsets under each code change. We used Python scripts complemented with manual analysis to query, compute, and label these factors. \black{Data from three repetitive experiments were then aggregated for model fitting, resulting in 1,263 responses from \textit{\black{GPT-4 (ChatGPT)}+\(\textbf{P}_{cid}\)} and 1,509 responses from \textit{DeepSeek-R1+\(\textbf{P}_{cot-guardrail}\)}. All continuous explanatory variables were standardized prior to regression analysis.}

\begin{figure*}[htbp]
    \centering
    \begin{minipage}[b]{0.45\linewidth}
        \centering
        \includegraphics[width=\linewidth]{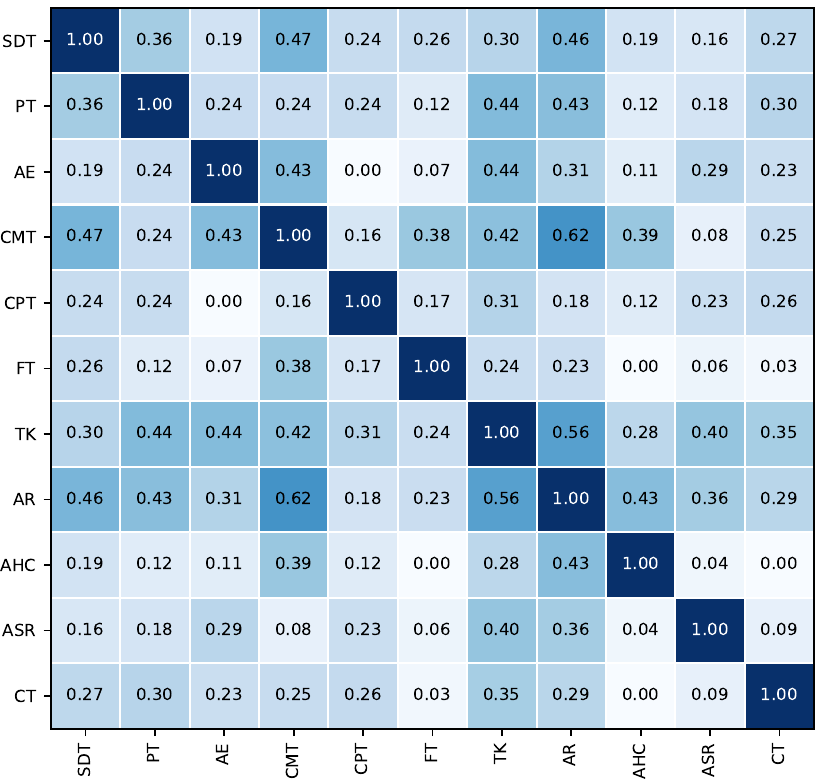}
        \par 
        \scriptsize DeepSeek-R1 
    \end{minipage}
    \hfill
    \begin{minipage}[b]{0.45\linewidth}
        \centering
        \includegraphics[width=\linewidth]{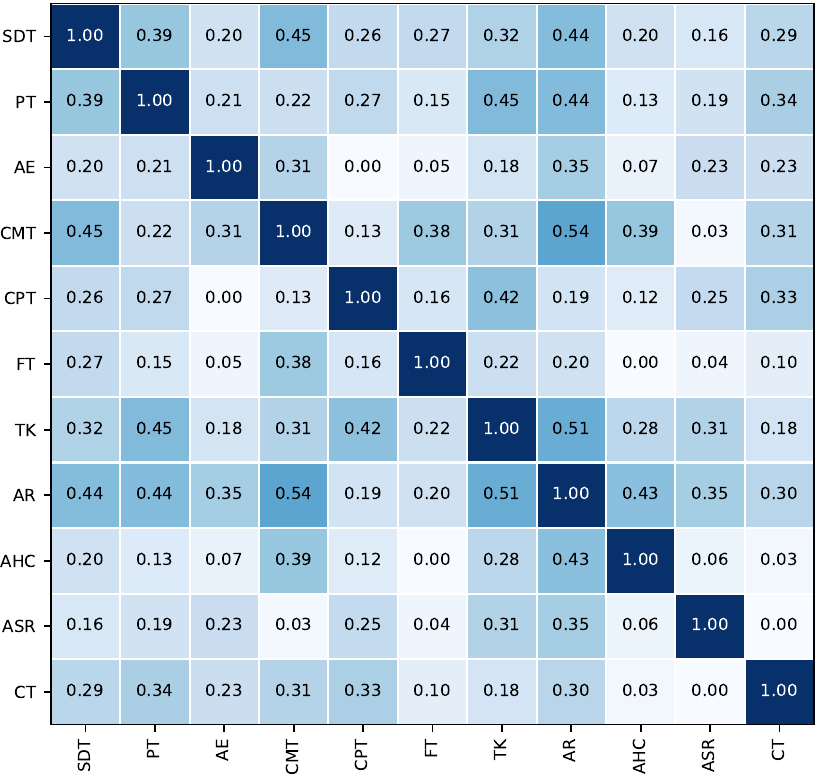}
        \par 
        \scriptsize GPT-4 (ChatGPT)
    \end{minipage}
    \caption{The Phi\_K correlation coefficients of explanatory variables}
    \label{fig:correlation}
\end{figure*}

\subsubsection{Regression Model Construction}
\mbox{}\\
\textbf{Correlation and Redundancy Analysis:} If explanatory variables are highly correlated, it can cause interference in model construction and analysis. As the predefined explanatory variables in our study include both categorical and continuous variables, we utilized a refined Pearson's hypothesis test of independence --- Phi\_K~\citep{baak2020new}, to measure the correlation between each factor. \black{The correlation coefficients of explanatory variables for \textit{\black{GPT-4 (ChatGPT)}+\(\textbf{P}_{cid}\)} and \textit{DeepSeek-R1+\(\textbf{P}_{cot-guardrail}\)} are illustrated in Fig.~\ref{fig:correlation}.} We chose $|\rho| > 0.7$ as the threshold because it is recommended as the threshold for strong correlation~\citep{hinkle2003applied}. From Fig.~\ref{fig:correlation}, where an abbreviation denotes each factor, it can be seen that the correlation coefficients between each pair of factors are all below 0.7, suggesting that there is no strong correlation between these factors. However, two factors without high correlation can still be duplicates, distorting the relationship between explanatory and response variables in model fitting. We encoded categorical variables as dummy variables for redundancy analysis~\citep{legendre2012numerical}. Utilizing the \textit{redun} function in the \texttt{rms} package\footnote{https://cran.r-project.org/web/packages/rms/index.html}, we assessed potential redundant variables with its default threshold of $R^2 \geq 0.9$, ultimately finding none.

\begin{figure*}[htbp]
    \centering
    \begin{minipage}[b]{0.45\linewidth}
        \centering
        \includegraphics[width=\linewidth]{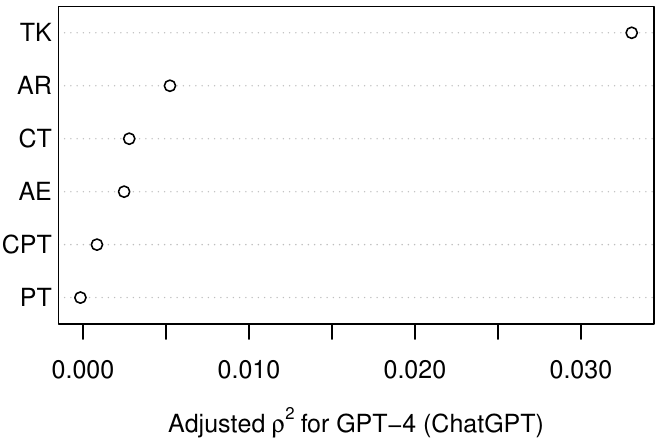}
    \end{minipage}
    \hfill
    \begin{minipage}[b]{0.45\linewidth}
        \centering
        \includegraphics[width=\linewidth]{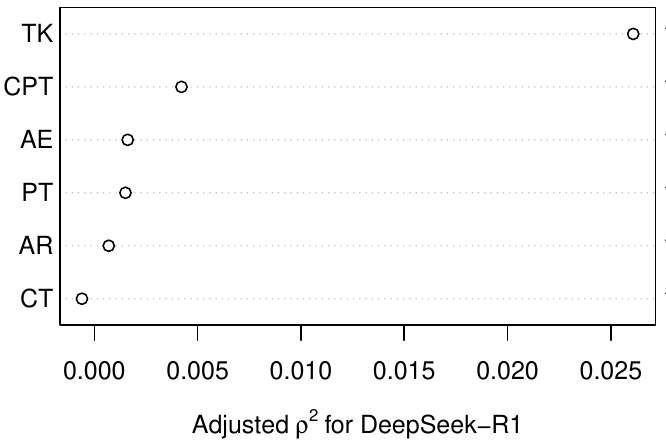}
    \end{minipage}
    \caption{Dotplot of the Spearman ${\rho}^2$ between explanatory and response variables across three repetitive experiments}
    \label{fig:spearman2}
\end{figure*}

\textbf{Degrees of Freedom Allocation:} \black{Overfitting often occurs when a model has too many free parameters to estimate for the amount of information in the data. To minimize this risk, we estimated the budget for degrees of freedom before assigning them to explanatory variables. As proposed by ~\cite{harrell2001regression}, we define the degrees of freedom budget as $\frac{1}{15}(n - \frac{1}{n^2} \sum_{i=1}^{k} n_i^3)$, where $n$ denotes the total sample size and $n_i$ represents the sample size of the $i$-th category of the ordinal response variable. Applying this formula yielded a budget of 79 for \textit{DeepSeek-R1+\(\textbf{P}_{cot-guardrail}\)} and 70 for \textit{\black{GPT-4 (ChatGPT)}+\(\textbf{P}_{cid}\)}. We calculated Spearman's rank correlations (${\rho}^2$) between each continuous explanatory variable and the dependent variable to measure the potential of nonlinear relationships between them. As suggested by ~\cite{harrell2001regression}, explanatory variables with larger ${\rho}^2$ can be assigned higher degrees of freedom, and the maximum allocated degrees of freedom for each variable should be limited to five to minimize the risk of overfitting.} Therefore, according to the computation results shown in Fig.~\ref{fig:spearman2}, \black{for both \textit{DeepSeek-R1+\(\textbf{P}_{cot-guardrail}\)} and \textit{GPT-4 (ChatGPT)+\(\textbf{P}_{cid}\)},} we utilized the \texttt{rcs} function from the R's \texttt{rms} package to allocate 3 degrees of freedom for the factor `Token' with the highest ${\rho}^2$, while the other factors were allocated 1 degree of freedom. Category variables were converted into factor variables, and their corresponding degrees of freedom were automatically allocated during model fitting.

\textbf{Model Fitting:} \black{Next, we fitted separate cumulative link models for \textit{GPT-4 (ChatGPT)+\(\textbf{P}_{cid}\)} and \textit{DeepSeek-R1+\(\textbf{P}_{cot-guardrail}\)} using the \black{\textit{clm}} function from the \textit{ordinal} R package~\citep{christensen2015package}. Both models exhibited several insignificant predictors. Given that data utilized in fitting spans multiple types of security defects, null effects at the aggregate level may hide significant type-specific effects. Therefore, we tested potential interactions between `SecurityDefectType' and other explanatory variables. Among the 10 factors excluding `SecurityDefectType', we hypothesized that the effects of `Token', `Position' and `Complexity' on LLM performance could vary across different security defect types. Specifically, for defects that can be identified based on local syntactic patterns, the effects of `Token', `Position' and `Complexity' may be limited. However, for defects requiring inter-procedural reasoning for detection, larger code size, higher code complexity, and deeper defect positions may increase the comprehension burden of LLMs. Accordingly, we incorporated three interaction terms--`SecurityDefectType × Token', `SecurityDefectType × Position', and `SecurityDefectType × Complexity'--into the regression models for \textit{\black{GPT-4 (ChatGPT)}+\(\textbf{P}_{cid}\)} and \textit{DeepSeek-R1+\(\textbf{P}_{cot-guardrail}\)}, and compared them against their counterparts without interactions. For \textit{DeepSeek-R1+\(\textbf{P}_{cot-guardrail}\)}, the `SecurityDefectType × Complexity' term was statistically significant ($p=0.02$), and the model fit improved as indicated by a reduction in its Akaike Information Criterion (AIC) from 2969.80 to 2959.46. Therefore, all interactions were retained in the regression model for DeepSeek-R1. For \textit{\black{GPT-4 (ChatGPT)}+\(\textbf{P}_{cid}\)}, none of the interaction terms were statistically significant and their inclusion led to a worse model fit (AIC increased from 2644.22 to 2655.60). Hence, all interactions were dropped from the model for GPT-4 (ChatGPT).} 


\textbf{Power Analysis:} \black{Statistical power for evaluation of influential factors was assessed via Monte Carlo simulations for ordinal regression models~\citep{gambarota2024ordinal}. Following Cohen's effect size guidelines~\citep{cohen2013statistical}, we selected small ($d=0.2$), medium ($d=0.5$), and large ($d=0.8$) effects and converted them into odds ratios of 1.43, 2.47, and 4.26, respectively. We calculated the statistical power for each considered effect size and each predictor, with an N of 1,263 for \textit{\black{GPT-4 (ChatGPT)}+\(\textbf{P}_{cid}\)} and 1,509 for \textit{DeepSeek-R1+\(\textbf{P}_{cot-guardrail}\)}. The resulting power estimates are presented in Table~\ref{tbl:power}.}
    
\black{All the results and scripts utilized in API-based LLM invocation, data labeling, problem extraction, and regression analysis are provided in our replication package~\citep{replpack}.}

\section{Results} \label{sec:result}
\subsection{Performance of LLMs (RQ1)}
\label{sec: rq1_result}
Based on the evaluation metrics defined in Section~\ref{sec:metrics}, we evaluated the capability of seven popular LLMs in security code review across different prompts. We compared the performance of LLMs with baseline tools in the C/C++ and Python datasets, respectively. If two of the three metrics were superior (i.e., higher I-Score or IH-Score, lower M-Score), we deemed that the current combination outperformed the others. 
Due to differences in the token limit of each LLM and the inherent token count of each prompt template, the number of responses successfully generated by each LLM-prompt combination also varied. To avoid bias due to different samples and to ensure rigorous comparisons, we also provided analyses on the intersection of source code files that all LLM-prompt combinations could generate responses, which yielded 57 instances. As a result, two sets of scores are provided: (1) the response count and performance scores of each LLM-prompt combination on the complete Python and C/C++ dataset (see Table~\ref{tbl:performance_python} and Table~\ref{tbl:performance_c}), and (2) the performance scores of each LLM-prompt combination on the selected 57 instances (see Table~\ref{tbl:performance_57_instances}). The best values for each score are highlighted in bold and marked with asterisks in the tables. 
It is evident that, DeepSeek-R1 outperformed all other models, followed by GPT-4 \black{provided in the ChatGPT platform}. Specifically, Table~\ref{tbl:performance_python} shows that on the Python dataset, the combination of DeepSeek-R1 with \(\textbf{P}_{cid}\) achieved the best performance. According to Table~\ref{tbl:performance_c}, on the C/C++ dataset, the combination of DeepSeek-R1 with \(\textbf{P}_{cot-guardrail}\) performed the best. Furthermore, Table~\ref{tbl:performance_57_instances} demonstrates that in the selected 57 cases, DeepSeek-R1 with \(\textbf{P}_{b}\) achieved better results than all other LLM-prompt combinations. Although \black{GPT-4 (ChatGPT)} did not perform as well as DeepSeek-R1, it still obviously surpassed the remaining LLMs. For the two best-performing LLMs, i.e, DeepSeek-R1 and \black{GPT-4 (ChatGPT)}, we calculated the performance scores of them across different prompt templates on the full dataset (including both Python and C/C++ datasets), as shown in Table~\ref{tbl:performance_gpt4_ds1}. We found that DeepSeek-R1 performed best with \(\textbf{P}_{cot-guardrail}\), whereas \black{GPT-4 (ChatGPT)} achieved its highest scores with \(\textbf{P}_{cid}\). On the whole, the capabilities of currently popular LLMs in conducting security code reviews are still limited. Notably, with enhanced prompts, the best performance of each LLM in security code review is significantly superior to that of static analysis tools. This could be attributed to two reasons: (1) for context-sensitive tools such as CodeQL, static analysis can be impacted by the lack of code context in our dataset~\citep{russo2010dynamic}; (2) there is a relatively high frequency of security defects related to multi-threading and asynchronous programming, such as race conditions. These defects often involve complex runtime behaviors that are better captured through dynamic analysis, and are therefore challenging for static analysis tools to detect~\citep{charoenwet2024empirical}. 

\begin{table*}[ht]
    \centering
    \begin{threeparttable}
    \scriptsize
    \caption{Performance scores and entropy for each LLM-Prompt combination on the Python dataset}
        \begin{tabular}{|p{2.2cm}|p{2cm}|p{1.5cm}|p{1.5cm}|p{1.5cm}|p{1.5cm}|p{1.5cm}|p{1.5cm}|}
        \hline
        \multicolumn{8}{|c|}{\textbf{Python}}  \\\hline
        \multicolumn{2}{|c|}{\textbf{Model \& Prompt}} & \textbf{\(\textbf{P}_b\)}   &\textbf{\(\textbf{P}_r\)}   &\textbf{\(\textbf{P}_c\)}   & \textbf{\(\textbf{P}_{cid}\)} &\textbf{\(\textbf{P}_{cot}\)} &\textbf{\(\textbf{P}_{cot-guardrail}\)}\\\hline
        
        
        \multirow{5}{*}{\textbf{GPT-4 (ChatGPT)}} & \textbf{Resp. Count} & \cellcolor{ncolor!17.5}175& \cellcolor{ncolor!17.5}175& \cellcolor{ncolor!17.5}175& \cellcolor{ncolor!17.4}174& \cellcolor{ncolor!17.4}174 &\\\cline{2-8}
        ~ & \textbf{I-Score} & \cellcolor{icolor!23.8}4.76\%& \cellcolor{icolor!17.15}3.43\%& \cellcolor{icolor!16.2}3.24\%& \cellcolor{icolor!25.85}5.17\%& \cellcolor{icolor!15.3}3.06\%&\\\cline{2-8}
        ~ & \textbf{IH-Score} & \cellcolor{ihcolor!46.7}9.34\%& \cellcolor{ihcolor!47.6}9.52\%& \cellcolor{ihcolor!41.9}8.38\%& \cellcolor{ihcolor!72.8}\textbf{14.56*}\%& \cellcolor{ihcolor!43.1}8.62\% &\\\cline{2-8}
        ~ & \textbf{M-Score} & \cellcolor{mcolor!38.69}61.31\%& \cellcolor{mcolor!39.05}60.95\%& \cellcolor{mcolor!53.53}46.47\%& \cellcolor{mcolor!55.94}44.06\%& \cellcolor{mcolor!38.31}61.69\% &\\\cline{2-8}
        ~ & \textbf{Entropy} & \cellcolor{ecolor!12.87}0.7426& \cellcolor{ecolor!13.59}0.7283& \cellcolor{ecolor!7}0.8600& \cellcolor{ecolor!2.92}0.9416&\cellcolor{ecolor!12.75}0.7450 &\\\hline

        \multirow{4}{*}{\textbf{GPT-4 (API)}} & \textbf{Resp. Count} &\cellcolor{ncolor!12.1}121 &\cellcolor{ncolor!12.1}121 &\cellcolor{ncolor!12.1}121 &\cellcolor{ncolor!12.1}119 &\cellcolor{ncolor!12.1}118&\\\cline{2-8}
        ~ &\textbf{I-Score} &\cellcolor{icolor!6.9}1.38\% &\cellcolor{icolor!1.4}0.28\% &\cellcolor{icolor!9.65}1.93\% &\cellcolor{icolor!8.4}1.68\% &\cellcolor{icolor!1.4}0.28\% &\\\cline{2-8}
        ~ &\textbf{IH-Score} &\cellcolor{ihcolor!16.55}3.31\% &\cellcolor{ihcolor!1.4}0.28\% &\cellcolor{ihcolor!13.75}2.75\% &\cellcolor{ihcolor!19.6}3.92\% &\cellcolor{ihcolor!1.4}0.28\% &\\\cline{2-8}
        ~ &\textbf{M-Score} &\cellcolor{mcolor!10.47}89.53\% &\cellcolor{mcolor!1.38}98.62\% &\cellcolor{mcolor!20.66}79.34\% &\cellcolor{mcolor!17.93}82.07\% &\cellcolor{mcolor!1.41}98.59\% &\\\cline{2-8}
        ~ &\textbf{Entropy} &\cellcolor{ecolor!45.45}0.0911 &\cellcolor{ecolor!48.11}0.0379 &\cellcolor{ecolor!36.17}0.2766 &\cellcolor{ecolor!35.76}0.2848 &\cellcolor{ecolor!49.2}\textbf{0.0160*} &\\\hline

        \multirow{5}{*}{\textbf{GPT-4 Turbo}} & \textbf{Resp. Count} & \cellcolor{ncolor!20.1}201& \cellcolor{ncolor!20.1}201& \cellcolor{ncolor!20.1}201& \cellcolor{ncolor!20.1}201& \cellcolor{ncolor!20.1}201 &\\\cline{2-8}        
        ~ & \textbf{I-Score} & \cellcolor{icolor!3.3}0.66\%&\cellcolor{icolor!3.3}1.65\% &\cellcolor{icolor!3.3}0.66 & \cellcolor{icolor!4.15}0.83 &  \cellcolor{icolor!0.65}0.17 &\\\cline{2-8}     
        ~ & \textbf{IH-Score} & \cellcolor{ihcolor!3.3}0.66 &\cellcolor{ihcolor!4.15}0.83 &\cellcolor{ihcolor!11.6}2.32 &\cellcolor{ihcolor!9.95}1.99 & \cellcolor{ihcolor!1.65}0.33 &\\\cline{2-8}
        ~ & \textbf{M-Score} & \cellcolor{mcolor!2.82}97.18\% &\cellcolor{mcolor!3.65}96.35\% &\cellcolor{mcolor!14.26}85.74\% &\cellcolor{mcolor!7.13}92.87\% &\cellcolor{mcolor!1.99}98.01\% &  \\\cline{2-8}
        ~ & \textbf{Entropy} & \cellcolor{ecolor!46.12}0.0777& \cellcolor{ecolor!45.50}0.0901& \cellcolor{ecolor!37.88}0.2424& \cellcolor{ecolor!40.64}0.1873&\cellcolor{ecolor!47.26}0.0548 &\\\hline

        \multirow{5}{*}{\textbf{Gemini Pro}} & \textbf{Resp. Count} & \cellcolor{ncolor!16.2}162& \cellcolor{ncolor!16.1}161& \cellcolor{ncolor!16.0}160& \cellcolor{ncolor!16.0}160& \cellcolor{ncolor!15.9}159 &\\\cline{2-8}    
        ~ & \textbf{I-Score} & \cellcolor{icolor!0.0}0.00\%& \cellcolor{icolor!0.0}0.00\%& \cellcolor{icolor!3.15}0.63\%& \cellcolor{icolor!6.25}1.25\%& \cellcolor{icolor!1.05}0.21\% &\\\cline{2-8}      
        ~ & \textbf{IH-Score} & \cellcolor{ihcolor!2.05}0.41\%& \cellcolor{ihcolor!0.0}0.00\%& \cellcolor{ihcolor!4.2}0.84\%& \cellcolor{ihcolor!17.7}3.54\%& \cellcolor{ihcolor!1.05}0.21\% &\\\cline{2-8}  
        ~ & \textbf{M-Score} & \cellcolor{mcolor!9.46}90.54\%& \cellcolor{mcolor!2.91}97.09\%& \cellcolor{mcolor!15.96}84.04\%& \cellcolor{mcolor!44.79}55.21\%& \cellcolor{mcolor!1.47}98.53\% &\\\cline{2-8}
        ~ & \textbf{Entropy} & \cellcolor{ecolor!41.22}0.1757& \cellcolor{ecolor!46.27}0.0746& \cellcolor{ecolor!33.15}0.3370& \cellcolor{ecolor!19.37}0.6125&\cellcolor{ecolor!47.98}0.0404 &\\\hline

        \multirow{5}{*}{\textbf{Llama 2 7B}} & \textbf{Resp. Count} & \cellcolor{ncolor!6.2}62& \cellcolor{ncolor!6.2}62& \cellcolor{ncolor!6.1}61& \cellcolor{ncolor!6.1}61&\cellcolor{ncolor!5.0}50 &\\\cline{2-8}
        ~ & \textbf{I-Score} & \cellcolor{icolor!0.0}0.00\%& \cellcolor{icolor!2.7}0.54\%& \cellcolor{icolor!8.1}1.62\%& \cellcolor{icolor!10.95}2.19\%&\cellcolor{icolor!6.665}1.33 &\\\cline{2-8}
        ~ & \textbf{IH-Score}& \cellcolor{ihcolor!8.05}1.61\%& \cellcolor{ihcolor!8.05}1.61\%& \cellcolor{ihcolor!27}5.40\%& \cellcolor{ihcolor!43.7}8.74\%&\cellcolor{ihcolor!23.35}4.67\% &\\\cline{2-8}
        ~ & \textbf{M-Score} & \cellcolor{mcolor!12.90}87.10\%& \cellcolor{mcolor!16.67}83.33\%& \cellcolor{mcolor!43.22}56.78\%& \cellcolor{mcolor!59.02}40.98\%&\cellcolor{mcolor!66.67}33.33\% &\\\cline{2-8}
        ~ & \textbf{Entropy} & \cellcolor{ecolor!36.88}0.2625& \cellcolor{ecolor!34.65}0.3070& \cellcolor{ecolor!26.17}0.4766& \cellcolor{ecolor!28.25}0.4351&\cellcolor{ecolor!22.40}0.5521 &\\\hline

        \multirow{5}{*}{\textbf{Llama 2 70B}} & \textbf{Resp. Count} & \cellcolor{ncolor!6.2}62& \cellcolor{ncolor!6.2}62& \cellcolor{ncolor!6.2}62& \cellcolor{ncolor!6.1}61&\cellcolor{ncolor!4.7}47 &\\\cline{2-8}\
        ~ & \textbf{I-Score} & \cellcolor{icolor!2.7}0.54\%& \cellcolor{icolor!5.35}1.07\%& \cellcolor{icolor!10.75}2.15\%& \cellcolor{icolor!19.15}3.83\%&\cellcolor{icolor!7.1}1.42\% &\\\cline{2-8}
        ~ & \textbf{IH-Score} & \cellcolor{ihcolor!24.2}4.84\%& \cellcolor{ihcolor!21.5}4.30\%& \cellcolor{ihcolor!32.25}6.45\%& \cellcolor{ihcolor!57.4}11.48\%&\cellcolor{ihcolor!17.75}3.55\% &\\\cline{2-8}
        ~ & \textbf{M-Score} & \cellcolor{mcolor!20.43}79.57\%& \cellcolor{mcolor!15.59}84.41\%& \cellcolor{mcolor!41.4}58.60\%& \cellcolor{mcolor!56.83}43.17\%&\cellcolor{mcolor!63.83}36.17\% &\\\cline{2-8}
        ~ & \textbf{Entropy} & \cellcolor{ecolor!27.65}0.4470& \cellcolor{ecolor!32.63}0.3474& \cellcolor{ecolor!32.43}0.3514& \cellcolor{ecolor!30.84}0.3832&\cellcolor{ecolor!21.19}0.5762 &\\\hline

        \multirow{5}{*}{\textbf{DeepSeek-R1}} & \textbf{Resp. Count} & \cellcolor{ncolor!23.4}234& \cellcolor{ncolor!23.4}234& \cellcolor{ncolor!23.4}234& \cellcolor{ncolor!23.4}234&\cellcolor{ncolor!23.4}234 &\cellcolor{ncolor!23.4}234\\\cline{2-8}\
        ~ & \textbf{I-Score} & \cellcolor{icolor!27.05}5.41\%& \cellcolor{icolor!31.35}6.27\%& \cellcolor{icolor!23.5}4.70\%& \cellcolor{icolor!36.3}\textbf{7.26\%*}&\cellcolor{icolor!22.8}4.56\% &\cellcolor{icolor!29.2}5.84\%\\\cline{2-8}
        ~ & \textbf{IH-Score} & \cellcolor{ihcolor!39.15}7.83\%& \cellcolor{ihcolor!37.75}7.55\%& \cellcolor{ihcolor!31.35}6.27\%& \cellcolor{ihcolor!46.3}9.26\%&\cellcolor{ihcolor!32.05}6.41\% &\cellcolor{ihcolor!43.45}8.69\%\\\cline{2-8}
        ~ & \textbf{M-Score} & \cellcolor{mcolor!63.96}36.04\% & \cellcolor{mcolor!67.95}32.05\%& \cellcolor{mcolor!70.66}\textbf{29.34\%*}&\cellcolor{mcolor!69.23}30.77\%&\cellcolor{mcolor!53.13}46.87\% &\cellcolor{mcolor!64.31}35.69\%\\\cline{2-8}
        ~ & \textbf{Entropy} & \cellcolor{ecolor!22.8}0.5440& \cellcolor{ecolor!23.64}0.5273& \cellcolor{ecolor!24.51}0.5098 & \cellcolor{ecolor!22.78}0.5444&\cellcolor{ecolor!21.59}0.5683 &\cellcolor{ecolor!22.37}0.5526\\\hline

          \multirow{5}{*}{\textbf{Baselines}}  &\textbf{SAST} &\textbf{SonarQube} &\textbf{CodeQL} &\textbf{Bandit} &\textbf{Semgrep} & &\\\cline{2-8}
        ~ & \textbf{Resp. Count} & \cellcolor{ncolor!25.8}258& \cellcolor{ncolor!25.8}258& \cellcolor{ncolor!25.8}258& \cellcolor{ncolor!25.8}258& &\\\cline{2-8}
        ~ & \textbf{I-Score} & \cellcolor{icolor!0.0}0.00\%& \cellcolor{icolor!3.9}0.39\%& \cellcolor{icolor!0.0}0.00\%& \cellcolor{icolor!0.0}0.00\%& &\\\cline{2-8}
        ~ & \textbf{IH-Score} & \cellcolor{ihcolor!0.0}0.00\%& \cellcolor{ihcolor!5.81}1.16\%& \cellcolor{ihcolor!5.8}1.16\%& \cellcolor{ihcolor!0.0}0.00\%& &\\\cline{2-8}
        ~ & \textbf{M-Score} & \cellcolor{mcolor!8.52}91.47\%& \cellcolor{mcolor!17.44}82.56\%& \cellcolor{mcolor!21.71}78.29\%& \cellcolor{mcolor!9.69}90.31\%& &\\\hline

        \end{tabular}
\label{tbl:performance_python}
\end{threeparttable}
\end{table*}
\vspace*{-3mm}
\begin{table*}[ht]
    \scriptsize
    \begin{threeparttable}
    \caption{Performance scores and entropy for each LLM-Prompt combination on the C/C++ dataset}
        \begin{tabular}{|p{2.2cm}|p{2cm}|p{1.5cm}|p{1.5cm}|p{1.5cm}|p{1.5cm}|p{1.5cm}|p{1.5cm}|}
        \hline
        \multicolumn{8}{|c|}{\textbf{C/C++}} \\\hline
        \multicolumn{2}{|c|}{\textbf{Model \& Prompt}} & \textbf{\(\textbf{P}_b\)}   &\textbf{\(\textbf{P}_r\)}   &\textbf{\(\textbf{P}_c\)}   & \textbf{\(\textbf{P}_{cid}\)} &\textbf{\(\textbf{P}_{cot}\)} &\textbf{\(\textbf{P}_{cot-guardrail}\)}\\\hline
        

        \multirow{4}{*}{\textbf{GPT-4 (ChatGPT)}}& \textbf{Resp. Count} & \cellcolor{ncolor!24.7}247& \cellcolor{ncolor!24.7}247& \cellcolor{ncolor!24.7}247& \cellcolor{ncolor!24.7}247& \cellcolor{ncolor!24.7}247&\\\cline{2-8}
        ~ & \textbf{I-Score} & \cellcolor{icolor!22.25}4.45\%& \cellcolor{icolor!19.55}3.91\%& \cellcolor{icolor!19.55}3.91\%& \cellcolor{icolor!27}5.40\%& \cellcolor{icolor!26.3}5.26\%&\\\cline{2-8}
        ~ & \textbf{IH-Score} & \cellcolor{ihcolor!39.8}7.96\%& \cellcolor{ihcolor!40.05}8.01\%& \cellcolor{ihcolor!46.55}9.31\%& \cellcolor{ihcolor!58.7}11.74\%& \cellcolor{ihcolor!57.35}11.47\%&\\\cline{2-8}
        ~ & \textbf{M-Score} & \cellcolor{mcolor!44.53}55.47\%& \cellcolor{mcolor!35.76}64.24\%& \cellcolor{mcolor!55.47}44.53\%& \cellcolor{mcolor!61.00}39.00\%& \cellcolor{mcolor!56.41}43.59\%&\\\cline{2-8}
        ~ & \textbf{Entropy} & \cellcolor{ecolor!18.93}0.6215& \cellcolor{ecolor!23.36}0.5329& \cellcolor{ecolor!17.48}0.6505& \cellcolor{ecolor!15.23}0.6955& \cellcolor{ecolor!13.35}0.7330 &\\\hline

        \multirow{4}{*}{\textbf{GPT-4 (API)}} & \textbf{Resp. Count} &\cellcolor{ncolor!12.3}123 &\cellcolor{ncolor!12.3}123 &\cellcolor{ncolor!12.3}123 &\cellcolor{ncolor!12.1}121 &\cellcolor{ncolor!11.4}114 &\\\cline{2-8}
        ~ &\textbf{I-Score} &\cellcolor{icolor!5.40}1.08\% &\cellcolor{icolor!4.05}0.81\% &\cellcolor{icolor!9.5}1.90\% &\cellcolor{icolor!9.65}1.93\% &\cellcolor{icolor!4.4}0.88\% &\\\cline{2-8}
        ~ &\textbf{IH-Score} &\cellcolor{ihcolor!10.95}2.19\% &\cellcolor{ihcolor!4.05}0.81\% &\cellcolor{ihcolor!17.6}3.52\% &\cellcolor{ihcolor!15.5}3.03\% &\cellcolor{ihcolor!4.4}0.88\% &\\\cline{2-8}
        ~ &\textbf{M-Score} &\cellcolor{mcolor!13.01}86.99\% &\cellcolor{mcolor!1.63}98.37\% &\cellcolor{mcolor!23.58}76.42\% &\cellcolor{mcolor!24.24}75.76\% &\cellcolor{mcolor!0.58}99.42\% &\\\cline{2-8}
        ~ &\textbf{Entropy} &\cellcolor{ecolor!42.16}0.1568 &\cellcolor{ecolor!48.88}\textbf{0.0224*} &\cellcolor{ecolor!36.06}0.2789 &\cellcolor{ecolor!33.24}0.3353 &\cellcolor{ecolor!47.97}0.0406 &\\\hline

        \multirow{4}{*}{\textbf{GPT-4 Turbo}}& \textbf{Resp. Count} & \cellcolor{ncolor!26.9}269& \cellcolor{ncolor!26.9}269& \cellcolor{ncolor!26.9}269& \cellcolor{ncolor!26.9}269& \cellcolor{ncolor!26.9}269&\\\cline{2-8}
        ~ & \textbf{I-Score} & \cellcolor{icolor!1.25}0.25\%& \cellcolor{icolor!0.6}0.12\%& \cellcolor{icolor!1.25}0.25\%& \cellcolor{icolor!2.5}0.50\%& \cellcolor{icolor!2.5}0.25\%&\\\cline{2-8}
        ~ & \textbf{IH-Score} & \cellcolor{ihcolor!1.25}0.25\%& \cellcolor{ihcolor!1.85}0.37\%& \cellcolor{ihcolor!6.2}1.24\%& \cellcolor{ihcolor!8.05}1.61\%& \cellcolor{ihcolor!5.6}1.12\%&\\\cline{2-8}
        ~ & \textbf{M-Score} & \cellcolor{mcolor!1.98}98.02\%& \cellcolor{mcolor!1.36}98.64\%& \cellcolor{mcolor!8.92}91.08\%& \cellcolor{mcolor!6.57}93.43\%& \cellcolor{mcolor!3.72}96.28\%&\\\cline{2-8}
        ~ & \textbf{Entropy} & \cellcolor{ecolor!47.44}0.0512& \cellcolor{ecolor!48.34}0.0332& \cellcolor{ecolor!43.00}0.1401& \cellcolor{ecolor!42.12}0.1576& \cellcolor{ecolor!44.88}0.1024 &\\\hline

        \multirow{4}{*}{\textbf{Gemini Pro}} & \textbf{Resp. Count} & \cellcolor{ncolor!22.3}223& \cellcolor{ncolor!22.1}221& \cellcolor{ncolor!22.1}221& \cellcolor{ncolor!22.1}221& \cellcolor{ncolor!22.0}220&\\\cline{2-8}
        ~ & \textbf{I-Score} & \cellcolor{icolor!0.0}0.00\%& \cellcolor{icolor!0.75}0.15\%& \cellcolor{icolor!1.5}0.30\%& \cellcolor{icolor!8.3}1.66\%& \cellcolor{icolor!0.00}0.00\%&\\\cline{2-8}
        ~ & \textbf{IH-Score} & \cellcolor{ihcolor!2.25}0.45\%& \cellcolor{ihcolor!0.75}0.15\%& \cellcolor{ihcolor!4.50}0.90\%& \cellcolor{ihcolor!18.85}3.77\%& \cellcolor{ihcolor!2.25}0.45\%&\\\cline{2-8}  
        ~ & \textbf{M-Score} & \cellcolor{mcolor!7.03}92.97\%& \cellcolor{mcolor!1.81}98.19\%& \cellcolor{mcolor!10.56}89.44\%& \cellcolor{mcolor!38.64}61.54\%& \cellcolor{mcolor!0.91}99.09\%&\\\cline{2-8}
        ~ & \textbf{Entropy} & \cellcolor{ecolor!42.85}0.1430 & \cellcolor{ecolor!47.72}0.0457& \cellcolor{ecolor!37.54}0.2493& \cellcolor{ecolor!22.94}0.5413& \cellcolor{ecolor!48.75}0.0250 &\\\hline

        \multirow{4}{*}{\textbf{Llama 2 7B}} & \textbf{Resp. Count} & \cellcolor{ncolor!5.7}57& \cellcolor{ncolor!5.7}57& \cellcolor{ncolor!5.7}57& \cellcolor{ncolor!5.1}51& \cellcolor{ncolor!3.3}33&\\\cline{2-8}
        ~ & \textbf{I-Score} & \cellcolor{icolor!2.9}0.58\%& \cellcolor{icolor!0.00}0.00\%& \cellcolor{icolor!0.00}0.00\%& \cellcolor{icolor!13.05}2.61\%&\cellcolor{icolor!5.05}1.01\% &\\\cline{2-8}
        ~ & \textbf{IH-Score}& \cellcolor{ihcolor!11.7}2.34\%& \cellcolor{ihcolor!17.55}3.51\%& \cellcolor{ihcolor!29.25}5.85\%& \cellcolor{ihcolor!22.9}4.58\%&\cellcolor{ihcolor!35.35}7.07\% &\\\cline{2-8}
        ~ & \textbf{M-Score} & \cellcolor{mcolor!9.4}90.06\%& \cellcolor{mcolor!12.87}87.13\%& \cellcolor{mcolor!18.71}81.29\%& \cellcolor{mcolor!47.06}52.94\%& \cellcolor{mcolor!72.73}27.27\% &\\\cline{2-8}
        ~ & \textbf{Entropy} & \cellcolor{ecolor!36.75}0.2650& \cellcolor{ecolor!39.18}0.2167& \cellcolor{ecolor!39.17}0.2167& \cellcolor{ecolor!29.54}0.4092& \cellcolor{ecolor!20.23}0.5954 &\\\hline

        \multirow{4}{*}{\textbf{Llama 2 70B}} & \textbf{Resp. Count} & \cellcolor{ncolor!5.7}57& \cellcolor{ncolor!5.7}57& \cellcolor{ncolor!5.7}57& \cellcolor{ncolor!5.1}51& \cellcolor{ncolor!3.7}37&\\\cline{2-8}
        ~ & \textbf{I-Score} & \cellcolor{icolor!5.85}1.17\%& \cellcolor{icolor!5.85}1.17\%& \cellcolor{icolor!8.75}1.75\%& \cellcolor{icolor!32.7}6.54\%&\cellcolor{icolor!9.0}1.80\% &\\\cline{2-8}
        ~ & \textbf{IH-Score} & \cellcolor{ihcolor!23.4}4.68\%& \cellcolor{ihcolor!14.6}2.92\%& \cellcolor{ihcolor!17.55}3.51\%& \cellcolor{ihcolor!75.15}15.03\%&\cellcolor{ihcolor!31.55}6.31\% &\\\cline{2-8}
        ~ & \textbf{M-Score} & \cellcolor{mcolor!32.75}67.25\%& \cellcolor{mcolor!19.88}80.12\%& \cellcolor{mcolor!33.92}66.08\%& \cellcolor{mcolor!59.48}40.52\%&\cellcolor{mcolor!45.95}54.05\%  &\\\cline{2-8}
        ~ & \textbf{Entropy} & \cellcolor{ecolor!21.67}0.5667 & \cellcolor{ecolor!29.28}0.4145& \cellcolor{ecolor!30.53}0.3895& \cellcolor{ecolor!22.43}0.5515& \cellcolor{ecolor!19.24}0.6153 &\\\hline

        \multirow{4}{*}{\textbf{DeepSeek-R1}} & \textbf{Resp. Count} & \cellcolor{ncolor!26.9}269& \cellcolor{ncolor!26.9}269& \cellcolor{ncolor!26.9}269& \cellcolor{ncolor!26.9}269& \cellcolor{ncolor!26.9}269& \cellcolor{ncolor!26.9}269\\\cline{2-8}
        ~ & \textbf{I-Score} & \cellcolor{icolor!63.2}12.64\%& \cellcolor{icolor!42.15}8.43\%& \cellcolor{icolor!49.55}9.91\%& \cellcolor{icolor!53.05}10.61\%&\cellcolor{icolor!48.95}9.79\% &\cellcolor{icolor!64.45}\textbf{12.89\%*}\\\cline{2-8}
        ~ & \textbf{IH-Score} & \cellcolor{ihcolor!77.45}\textbf{15.49\%*}& \cellcolor{ihcolor!52.05}10.41\%& \cellcolor{ihcolor!55.75}11.15\%& \cellcolor{ihcolor!61.35}12.27\%&\cellcolor{ihcolor!63.2}12.64\% &\cellcolor{ihcolor!77.45}\textbf{15.49\%*}\\\cline{2-8}
        ~ & \textbf{M-Score} & \cellcolor{mcolor!70.01}29.99\%& \cellcolor{mcolor!79.06}\textbf{20.94\%*}& \cellcolor{mcolor!78.07}21.93\%& \cellcolor{mcolor!76.95}23.05\%&\cellcolor{mcolor!65.68}34.32\%  &\cellcolor{mcolor!72.61}27.39\\\cline{2-8}
        ~ & \textbf{Entropy} & \cellcolor{ecolor!22.95}0.5411 & \cellcolor{ecolor!25.33}0.4934& \cellcolor{ecolor!25.58}0.4884& \cellcolor{ecolor!24.98}0.5002& \cellcolor{ecolor!21.73}0.5654 &\cellcolor{ecolor!21.69}0.5663\\\hline

        \multirow{5}{*}{\textbf{Baselines}} &\textbf{SAST}  &\textbf{CppCheck} &\textbf{Semgrep} &\multicolumn{3}{c|}{}  &\\\cline{2-8}
        ~ & \textbf{Resp. Count} & \cellcolor{ncolor!27.6}276& \cellcolor{ncolor!27.6}276&\multicolumn{3}{c|}{}&\\\cline{2-8}
        ~ & \textbf{I-Score} & \cellcolor{icolor!0.0}0.00\%& \cellcolor{icolor!0.0}0.00\%&\multicolumn{3}{c|}{} &\\\cline{2-8}
        ~ & \textbf{IH-Score} & \cellcolor{ihcolor!3.62}0.72\%& \cellcolor{ihcolor!1.81}0.36\%&\multicolumn{3}{c|}{} &\\\cline{2-8}
        ~ & \textbf{M-Score} & \cellcolor{mcolor!1.81}98.19\%& \cellcolor{mcolor!3.26}96.74\%&\multicolumn{3}{c|}{}  &\\\hline
         
        \end{tabular}
\label{tbl:performance_c}
\end{threeparttable}
\end{table*}
\begin{table*}
\centering
\scriptsize
\caption{Performance scores for each LLM-Prompt combination on 57 instances that all LLM-prompt combinations could generate responses}
\begin{tabular}{|p{2.2cm}|p{2cm}|p{1.5cm}|p{1.5cm}|p{1.5cm}|p{1.5cm}|p{1.5cm}|p{1.5cm}|}
\hline
\multicolumn{2}{|c|}{\textbf{Model \& Prompt}} & \textbf{\(\textbf{P}_b\)}   &\textbf{\(\textbf{P}_r\)}   &\textbf{\(\textbf{P}_c\)}   & \textbf{\(\textbf{P}_{cid}\)} &\textbf{\(\textbf{P}_{cot}\)}  &\textbf{\(\textbf{P}_{cot}\)}\\\hline


\multirow{3}{*}{\textbf{GPT-4 (ChatGPT)}} & \textbf{I-Score} & \cellcolor{icolor!43.85}8.77\%& \cellcolor{icolor!43.85}8.77\%& \cellcolor{icolor!29.25}5.85\%& \cellcolor{icolor!64.35}12.87\%& \cellcolor{icolor!49.7}9.94&\\\cline{2-8}
~ & \textbf{IH-Score}& \cellcolor{ihcolor!61.38}20.46\%& \cellcolor{ihcolor!57.9}19.30\%& \cellcolor{ihcolor!42.12}14.04\%& \cellcolor{ihcolor!71.94}\textbf{23.98*}\%& \cellcolor{ihcolor!59.64}19.88&\\\cline{2-8}
~ & \textbf{M-Score} & \cellcolor{mcolor!56.14}43.86\%& \cellcolor{mcolor!57.89}42.11\%& \cellcolor{mcolor!54.39}45.61\%& \cellcolor{mcolor!59.65}40.35\%& \cellcolor{mcolor!58.48}41.52\%&\\\hline

\multirow{3}{*}{\textbf{GPT-4 (API)}} & \textbf{I-Score} & \cellcolor{icolor!2.9}0.58\%& \cellcolor{icolor!0}0.00\%& \cellcolor{icolor!11.7}2.34\%& \cellcolor{icolor!17.55}3.51\%& \cellcolor{icolor!0}0.00&\\\cline{2-8}
~ & \textbf{IH-Score}& \cellcolor{ihcolor!20.45}4.09\%& \cellcolor{ihcolor!0}0\%& \cellcolor{ihcolor!23.4}4.68\%& \cellcolor{ihcolor!32.15}6.43\%& \cellcolor{ihcolor!2.9}0.58&\\\cline{2-8}
~ & \textbf{M-Score} & \cellcolor{mcolor!15.2}84.80\%& \cellcolor{mcolor!2.34}97.66\%& \cellcolor{mcolor!25.15}74.85\%& \cellcolor{mcolor!23.98}76.02\%& \cellcolor{mcolor!0.58}99.42\%&\\\hline

\multirow{3}{*}{\textbf{GPT-4 Turbo}} & \textbf{I-Score} & \cellcolor{icolor!2.9}0.58\% & \cellcolor{icolor!8.75}1.75\% & \cellcolor{icolor!11.7}2.34\%& \cellcolor{icolor!11.7}2.34\%& \cellcolor{icolor!0.00}0.00\%&\\\cline{2-8}
~ & \textbf{IH-Score}& \cellcolor{ihcolor!1.74}0.58\% & \cellcolor{ihcolor!10.53}3.51\%& \cellcolor{ihcolor!15.78}5.26\%& \cellcolor{ihcolor!12.27}4.09\%& \cellcolor{ihcolor!5.25}1.75\%&\\\cline{2-8}
~ & \textbf{M-Score} & \cellcolor{mcolor!2.92}97.08\%& \cellcolor{mcolor!8.77}91.23\%& \cellcolor{mcolor!24.56}75.44\%& \cellcolor{mcolor!11.7}88.30\%& \cellcolor{mcolor!7.02}92.98\%&\\\hline

\multirow{3}{*}{\textbf{Gemini Pro}} & \textbf{I-Score} & \cellcolor{icolor!2.9}0.58\% & 0.00\% & 0.00\%& \cellcolor{icolor!2.9}0.58\%& 0.00\%&\\\cline{2-8}
~ & \textbf{IH-Score}& \cellcolor{ihcolor!3.51}1.17\% & 0.00\% & \cellcolor{ihcolor!1.74}0.58\%& \cellcolor{ihcolor!8.76}2.92\%& 0.00\%&\\\cline{2-8}
~ & \textbf{M-Score} & \cellcolor{mcolor!3.51}96.49\%& 100.00\%& \cellcolor{mcolor!11.7}88.30\%& \cellcolor{mcolor!23.39}76.61\%& 100.00\%&\\\hline

\multirow{3}{*}{\textbf{Llama 2 7B}} & \textbf{I-Score} & \cellcolor{icolor!2.9}0.58\% & \cellcolor{icolor!5.85}1.17\%& \cellcolor{icolor!5.85}1.17\%& \cellcolor{icolor!2.9}0.58\%& \cellcolor{icolor!8.75}1.75\%&\\\cline{2-8}
~ & \textbf{IH-Score}& \cellcolor{ihcolor!5.25}1.75\%& \cellcolor{ihcolor!7.02}2.34\%& \cellcolor{ihcolor!12.27}4.09\%& \cellcolor{ihcolor!7.02}2.34\%&\cellcolor{ihcolor!22.8}7.60\%&\\\cline{2-8}
~ & \textbf{M-Score} & \cellcolor{mcolor!12.87}87.13\%& \cellcolor{mcolor!15.79}84.21\%& \cellcolor{mcolor!39.18}60.82\%& \cellcolor{mcolor!46.2}53.80\%& \cellcolor{mcolor!74.27}25.73\%&\\\hline

\multirow{3}{*}{\textbf{Llama 2 70B}} & \textbf{I-Score} & \cellcolor{icolor!11.7}2.34\%& \cellcolor{icolor!5.85}1.17\%& \cellcolor{icolor!14.6}2.92\%& \cellcolor{icolor!26.3}5.26\%& \cellcolor{icolor!20.3}4.09\%&\\\cline{2-8}
~ & \textbf{IH-Score}& \cellcolor{ihcolor!14.04}4.68\%& \cellcolor{ihcolor!7.02}2.34\%& \cellcolor{ihcolor!21.06}7.02\%& \cellcolor{ihcolor!52.62}17.54\%& \cellcolor{ihcolor!14.04}4.68\%&\\\cline{2-8}
~ & \textbf{M-Score} & \cellcolor{mcolor!29.82}70.18\%& \cellcolor{mcolor!19.88}80.12\%& \cellcolor{mcolor!57.31}42.69\%& \cellcolor{mcolor!79.53}\textbf{20.47*}\%& \cellcolor{mcolor!55.56}44.44\%&\\\hline

\multirow{3}{*}{\textbf{DeekSeek-R1}} & \textbf{I-Score} & \cellcolor{icolor!84.8}\textbf{16.96\%*}& \cellcolor{icolor!81.85}16.37\%& \cellcolor{icolor!64.35}12.87\%& \cellcolor{icolor!81.85}16.37\%& \cellcolor{icolor!67.25}13.45\%&\cellcolor{icolor!61.40}12.28\%\\\cline{2-8}
~ & \textbf{IH-Score}& \cellcolor{ihcolor!71.94}\textbf{23.98\%*}& \cellcolor{ihcolor!59.64}19.88\%& \cellcolor{ihcolor!94.74}15.79\%& \cellcolor{ihcolor!64.92}21.64\%& \cellcolor{ihcolor!56.13}18.71\%&\cellcolor{ihcolor!54.39}18.13\%\\\cline{2-8}
~ & \textbf{M-Score} & \cellcolor{mcolor!59.65}40.35\%& \cellcolor{mcolor!62.57}37.43\%& \cellcolor{mcolor!64.33}35.67\%& \cellcolor{mcolor!69.01}30.99\%& \cellcolor{mcolor!54.39}45.61\%&\cellcolor{mcolor!67.25}32.75\%\\\hline

\end{tabular}
\label{tbl:performance_57_instances}
\end{table*}
\begin{table*}[ht]
    \scriptsize
    \begin{threeparttable}
    \caption{Performance scores and entropy for GPT-4 and DeepSeek-R1 under different prompts on the complete dataset}
        \begin{tabular}{|p{2.2cm}|p{2cm}|p{1.5cm}|p{1.5cm}|p{1.5cm}|p{1.5cm}|p{1.5cm}|p{1.5cm}|}
        \hline
        \multicolumn{2}{|c|}{\textbf{Model \& Prompt}} & \textbf{\(\textbf{P}_b\)}   &\textbf{\(\textbf{P}_r\)}   &\textbf{\(\textbf{P}_c\)}   & \textbf{\(\textbf{P}_{cid}\)} &\textbf{\(\textbf{P}_{cot}\)} &\textbf{\(\textbf{P}_{cot-guardrail}\)}\\\hline
        \multirow{4}{*}{\textbf{GPT-4 (ChatGPT)}}& \textbf{Resp. Count} & \cellcolor{ncolor!21.1}422& \cellcolor{ncolor!21.1}422& \cellcolor{ncolor!21.1}422& \cellcolor{ncolor!21.05}421& \cellcolor{ncolor!21.05}421&\\\cline{2-8}
        ~ & \textbf{I-Score} & \cellcolor{icolor!22.9}4.58\%& \cellcolor{icolor!18.55}3.71\%& \cellcolor{icolor!18.15}3.63\%& \cellcolor{icolor!26.5}\textbf{5.30\%*}& \cellcolor{icolor!21.75}4.35\%&\\\cline{2-8}
        ~ & \textbf{IH-Score} & \cellcolor{ihcolor!42.65}8.53\%& \cellcolor{ihcolor!43.15}8.63\%& \cellcolor{ihcolor!44.6}8.92\%& \cellcolor{ihcolor!64.55}\textbf{12.91\%*}& \cellcolor{ihcolor!51.45}10.29\%&\\\cline{2-8}
        ~ & \textbf{M-Score} & \cellcolor{mcolor!42.11}57.89\%& \cellcolor{mcolor!37.12}62.88\%& \cellcolor{mcolor!54.67}45.33\%& \cellcolor{mcolor!58.91}\textbf{41.09\%*}& \cellcolor{mcolor!48.93}51.07\%&\\\hline

        \multirow{4}{*}{\textbf{DeepSeek-R1}} & \textbf{Resp. Count} & \cellcolor{ncolor!25.15}503& \cellcolor{ncolor!25.15}503& \cellcolor{ncolor!25.15}503& \cellcolor{ncolor!25.15}503& \cellcolor{ncolor!25.15}503& \cellcolor{ncolor!25.15}503\\\cline{2-8}
        ~ & \textbf{I-Score} & \cellcolor{icolor!46.4}9.28\%& \cellcolor{icolor!37.15}7.43\%& \cellcolor{icolor!37.45}7.49\%& \cellcolor{icolor!45.25}9.05\%&\cellcolor{icolor!36.8}7.36\% &\cellcolor{icolor!48.05}\textbf{9.61\%*}\\\cline{2-8}
        ~ & \textbf{IH-Score} & \cellcolor{ihcolor!59.65}11.93\%& \cellcolor{ihcolor!45.4}9.08\%& \cellcolor{ihcolor!44.4}8.88\%& \cellcolor{ihcolor!54.35}10.87\%&\cellcolor{ihcolor!48.7}9.74\% &\cellcolor{ihcolor!61.65}\textbf{12.33\%*}\\\cline{2-8}
        ~ & \textbf{M-Score} & \cellcolor{mcolor!67.2}32.80\%& \cellcolor{mcolor!73.89}26.11\%& \cellcolor{mcolor!74.62}\textbf{25.38\%*} & \cellcolor{mcolor!73.36}26.64\%&\cellcolor{mcolor!59.84}40.16\%  &\cellcolor{mcolor!68.75}31.25\\\hline

        \end{tabular}
\label{tbl:performance_gpt4_ds1}
\end{threeparttable}
\end{table*}

\subsubsection{RQ1.1}
When using basic prompts constructed by \(\textbf{P}_b\) (see Table~\ref{tbl:performance_python} and Table~\ref{tbl:performance_c}), DeepSeek-R1 performed the best, obtaining the highest I-Score (Python: 5.41\%, C/C++: 12.64\%), and the lowest M-Score (Python: 36.04\%, C/C++: 29.99\%). For the IH-Score, DeepSeek-R1 still achieved the highest value on the C/C++ dataset, reaching 15.49\%, while on the Python dataset, \black{GPT-4 accessed through ChatGPT} outperformed DeepSeek-R1 (7.83\%) with a score of 9.34\%. \black{GPT-4 (ChatGPT)} was regarded as the second-best model, followed by Llama 2 70B and \black{GPT-4 (API)}. GPT-4 Turbo, Gemini Pro, and Llama 2 7B performed the worst, with their performance in the Python dataset even inferior to that of the baseline tool, Bandit. Similar results are observed in Table~\ref{tbl:performance_57_instances}. 
Specifically, in the selected 57 instances, under the basic prompt \(\textbf{P}_b\), DeepSeek-R1 achieved I-Score = 16.96\%, IH-Score = 23.98\%, and M-Score = 40.35\%. It is evident that DeepSeek-R1's performance on the selected 57 instances exceeds its performance across the entire dataset. Given that these 57 code files are instances with relatively fewer tokens within our dataset, it could be speculated that the quantity of tokens input to LLM affects its performance in detecting security defects, which has been further explored and analyzed in RQ3.

\subsubsection{RQ1.2}
\textbf{\(\textbf{P}_r\):} As shown in Table~\ref{tbl:performance_python} and Table~\ref{tbl:performance_c}, the changes of the performance scores of each LLM under prompt \(\textbf{P}_r\) compared with that under \(\textbf{P}_b\) is generally small. Specifically, DeepSeek-R1, \black{GPT-4 (ChatGPT)}, and GPT-4 Turbo show subtle increases and decreases in performance on the Python and C/C++ datasets, respectively. Gemini Pro, Llama 2 70B \black{and GPT-4 (API)} demonstrate performance decreases on both datasets, while Llama 2 7B shows a slight improvement. A possible reason is that the name of components or packages utilized may already imply the source project of the code file, thereby diminishing the impact of project information provided in \(\textbf{P}_r\). Furthermore, the improvement of emphasizing the task through adding a persona in the prompt is LLM-specific. We speculate that some LLMs may already understand the task described in the prompt quite well, and thus, introducing a persona could add unnecessary complexity to the prompt and lead to distractions.

\textbf{\(\textbf{P}_c\):} We can see from Table~\ref{tbl:performance_python} and Table~\ref{tbl:performance_c} that compared to using the basic prompt \(\textbf{P}_b\), the performance of DeepSeek-R1 significantly decreases under \(\textbf{P}_c\) in both Python and C/C++ datasets. The performances of \black{GPT-4 (ChatGPT)} with \(\textbf{P}_c\) exhibit different changes across different datasets, with an improvement in the C/C++ dataset but a decrease in the Python dataset. However, all other LLMs achieved higher performance scores across both datasets. 

\textbf{\(\textbf{P}_{cid}\):} Under \(\textbf{P}_{cid}\), \black{GPT-4 (via both ChatGPT and API)}, GPT-4 Turbo, Gemini Pro, Llama 2 7B and Llama 2 70B all achieved their best performance on the Python and C/C++ datasets. However, \black{when applying \(\textbf{P}_{cid}\) to DeepSeek-R1, we observed a performance decline on the C/C++ dataset and an improvement on the Python dataset. After aggregating the results from both datasets, DeepSeek-R1 ultimately exhibited an overall decline in its performance metrics (see Table~\ref{tbl:performance_gpt4_ds1}). Compared with the results of \(\textbf{P}_c\), it can be inferred that incorporating specific CWE entries into the prompt is more effective than providing a generic instruction related to CWE. Moreover, the prompting strategy of incorporating CWE information may yield a more significant performance improvement on general-purpose LLMs than on the reasoning-optimized LLM in security code review.} We further analyzed why the impact of \(\textbf{P}_{cid}\) on the performance of DeepSeek-R1 \black{differs between the Python and C/C++ datasets}. As mentioned in Section~\ref{sec:dataset}, \textit{Race Condition} is the most prevalent security defect type in the Python dataset. The CWE list provided in \(\textbf{P}_{cid}\) includes CWE-435, which is highly relevant to \textit{Race Condition}, thereby enhancing model performance. In contrast, the C/C++ dataset features a more balanced distribution of security defect types, such as \textit{Integer overflow}, \textit{Resource Leak} and \textit{Crash}. We included solely high-level CWE entries in the CWE list and some of them, such as CWE-682 (Incorrect Calculation), do not exactly match certain security defect types like \textit{Integer Overflow}. These CWE entries failed to provide sufficient guidance and even introduced interference. More fine-grained subtypes may provide better guidance for specific security defect types. Therefore, determining the optimal granularity of CWE as auxiliary information for different LLMs in security defect detection warrants further investigation in future work.

\subsubsection{RQ1.3}
Table~\ref{tbl:performance_python} and Table~\ref{tbl:performance_c} illustrate that \(\textbf{P}_{cot}\) significantly enhances the ability to detect security defects for Llama 2 7B and Llama 2 70B in both Python and C/C++ datasets. However, for Gemini Pro, GPT-4 Turbo, DeepSeek-R1 \black{and GPT-4 (API)}, there is no improvement and even a detrimental effect to these LLMs. As for \black{GPT-4 (ChatGPT), this model exhibited variations in performance across different datasets.} It can be seen that the impact of supplementing missing context by CoT prompting on the performance of LLMs is multifaceted and unstable. A possible explanation is that different LLMs vary in their ability to process and understand input. Moreover, we instructed the LLM to generate missing code context based on the corresponding commit message in \(\textbf{P}_{cot}\). On the one hand, the generated context may contain the critical information needed to analyze the target security defect, which can help the LLM in security code review. On the other hand, the consistency between this generated context and the actual code context cannot be guaranteed. We observed that in some cases, the code context generated under Prompt \(\textbf{P}_{cot-1}\) was inconsistent with the actual code context, resulting in the original security defect no longer being identified as a defect within the generated context. Consequently, when Prompt \(\textbf{P}_{cot-2}\) was applied, the detection of security defects in the provided code was misled by the generated code context and yielded erroneous results.

To isolate the impact of LLM-generated code context on model performance, we designed \(\textbf{P}_{cot-guardrail}\) and conducted experiments under this prompt on the best-performing model, DeepSeek-R1. \(\textbf{P}_{cot-guardrail}\) retains CoT guidance and commit message but does not include the fabricated code context. The results are also presented in Table~\ref{tbl:performance_python}, Table~\ref{tbl:performance_c} and Table~\ref{tbl:performance_57_instances}. We observed that, on both the Python and C/C++ datasets, \(\textbf{P}_{cot-guardrail}\) consistently improved the performance of DeepSeek-R1 compared with \(\textbf{P}_{b}\), whereas \(\textbf{P}_{cot}\) led to a performance decline. This suggests that for DeepSeek-R1, LLM-generated code context introduced interference for security code review. However, in the selected 57 instances, DeepSeek-R1 performed worse with \(\textbf{P}_{cot-guardrail}\) than with \(\textbf{P}_{b}\). The 57 samples satisfy the token limits of all LLM-prompt combinations in our experiments, making them the samples with the fewest tokens in the dataset. Based on this, we hypothesize the following explanation: the commit message included in \(\textbf{P}_{cot-guardrail}\) helps the model understand code semantics and the CoT instruction drives the model to perform step-by-step reasoning, making this prompt more suitable for assisting the model in analyzing security defects in complex and long code. For code files with fewer tokens, these additional guidance could be unnecessary or even introduce noise to the LLM.

Missing code context can impact the reliability of model outcomes~\citep{li2025everything}, which remains a challenge when applying LLMs to real-world security code review. We attempted to generate code context using the LLM itself, but did not directly measure how often the LLM hallucinated irrelevant code and how such hallucinations affected model performance. This is because evaluating the consistency between LLM-generated code and the actual source code is challenging, as it requires a comprehensive understanding of the code repository and its historical commits. We plan to address this in future work. Some works have performed context augmentation by incorporating call graphs extracted through static analysis tools~\citep{luo2024fellmvp} or by utilizing caller-callee relationships within the code~\citep{wen2024vuleval, gao2023far, sun2024llm4vuln}. Given this, we will also compare these approaches with the LLM-driven code context generation in assisting LLMs with security code reviews, and explore how to better leverage the capabilities of LLMs to obtain more effective and precise code context.

\subsubsection{RQ1.4}
\begin{figure*}
  \centering
  \includegraphics[width=1\linewidth]{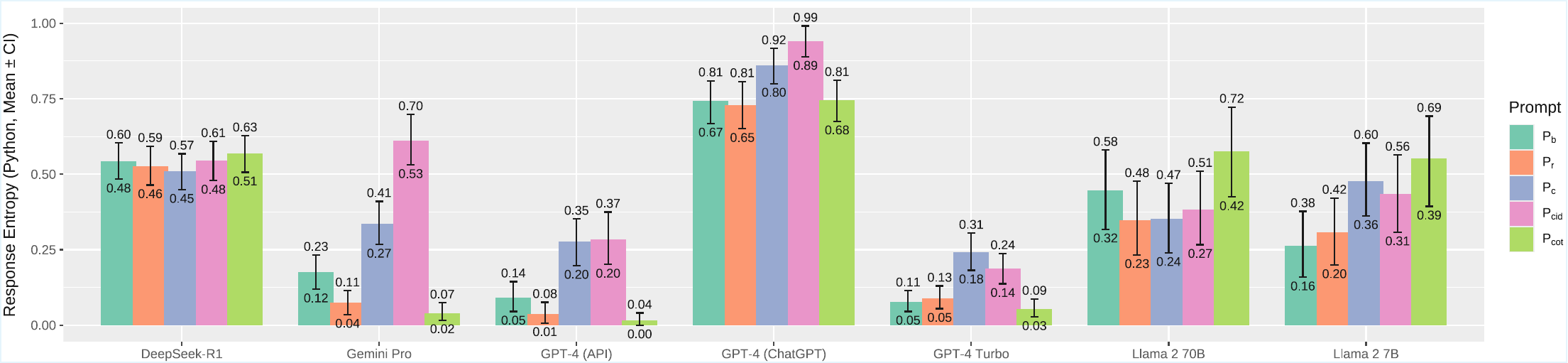}
  \caption{Response Entropy of each LLM-prompt combinations on the Python dataset}
  \label{fig:p_entropy}
  \vspace*{-3mm}
\end{figure*}

\begin{figure*}
  \centering
  \includegraphics[width=1\linewidth]{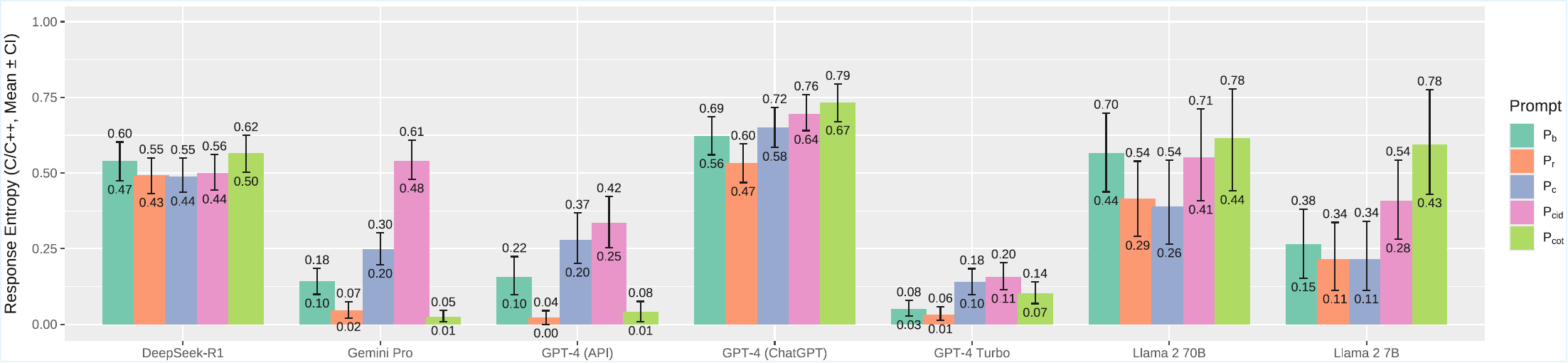}
  \caption{Response Entropy of each LLM-prompt combinations on the C/C++ dataset}
  \label{fig:c_entropy}
  \vspace*{-3mm}
\end{figure*}

According to the mean entropies and their 95\% confidence intervals (CI) of various LLM-prompt combinations in Fig.~\ref{fig:p_entropy} and Fig.~\ref{fig:c_entropy}, we found that GPT-4 Turbo, Gemini Pro and \black{GPT-4 (API)} exhibit relatively higher consistency in their responses across three repetitive experiments, followed by Llama 2 7B, Llama 2 70B and DeepSeek-R1, with \black{GPT-4 (ChatGPT)} showing the least consistency. 

Specifically, on the Python dataset, the highest consistency of responses was observed under \(\textbf{P}_{cot}\) with \black{GPT-4 (API), with a mean entropy of 0.0160 \black{(95\% CI: [0.00, 0.04])}. On the C/C++ dataset, the combination of GPT-4 (API) and \(\textbf{P}_r\) demonstrates the lowest mean entropy of 0.0224 \black{(95\% CI: [0.00, 0.04])}}. Given these two LLM-prompt combinations' poor security defect detection capabilities, their high consistency may be due to their tendency not to detect any security defects in the code. 
On the Python dataset, the highest average entropy of 0.9416 (95\% CI: [0.89, 0.99]) was observed with the combination of \black{GPT-4 (ChatGPT)} and \(\textbf{P}_{cot}\). On the C/C++ dataset, \black{GPT-4 (ChatGPT)} paired with \(\textbf{P}_{cid}\) exhibited the highest average entropy of 0.7330, with a 95\% CI from 0.67 to 0.79. Compared with other LLMs excluding DeepSeek-R1, \textbf{GPT-4 (ChatGPT)} not only demonstrated significantly superior performance, but also exhibited the highest uncertainty. In comparison with \textbf{GPT-4 (ChatGPT)}, DeepSeek-R1 showed improved performance in the security code review task but produced more stable outputs. However, the average entropy of DeepSeek-R1 remained relatively high among all models evaluated in our experiments. As shown in Table~\ref{tbl:llms}, we kept the decoding parameters at their default settings for all models. Except for DeepSeek-R1, the temperature and top\_p values of all LLMs fell within a similar range (0.9–1.0). Despite this, we observed substantial differences in response consistency across models. Furthermore, although DeepSeek-R1 was configured with a lower top\_p value (0.7), its response consistency was still relatively poor among all models. This observation suggests that decoding parameters alone may have a limited effect on regulating output consistency, and the observed differences are more likely attributable to model-intrinsic randomness.

\subsection{Quality Problems in Responses (RQ2)}
\label{sec: rq2_result}
\vspace{1em}
\subsubsection{RQ2.1}
\begin{figure*}
  \centering
  \includegraphics[width=1\linewidth]{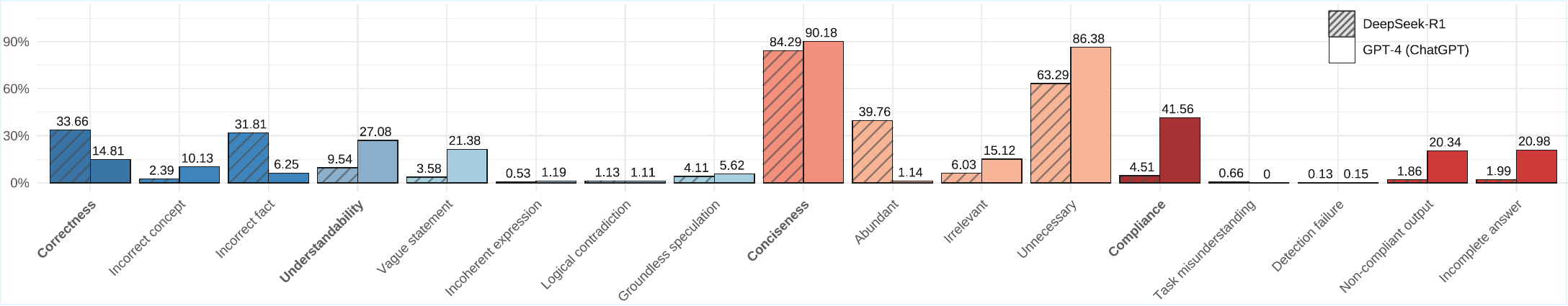}
  \caption{Average proportion of each problem type present in responses generated by GPT-4(ChatGPT) and DeepSeek-R1}
  \label{fig:problems_in_responses}
  \vspace*{-5mm}
\end{figure*}
The average distribution of 13 problem types across four themes in \black{the responses of \textit{\black{GPT-4 (ChatGPT)}+\(\textbf{P}_{cid}\)} and \textit{DeepSeek-R1+\(\textbf{P}_{cot-guardrail}\)}} to three repetitive experiments is illustrated in Fig.~\ref{fig:problems_in_responses}. \black{The same color represents types within the same theme, with darker shades indicating the corresponding theme. We differentiate the two LLM+prompt combinations by whether the bars are filled with stripe patterns. For brevity, we will refer to the two combinations as \black{GPT-4 (ChatGPT)} and DeepSeek-R1 in the rest of this section. The four themes are \textbf{\textit{Correctness}}, \textbf{\textit{Understandability}}, \textbf{\textit{Conciseness}} and \textbf{\textit{Compliance}}. \textbf{\textit{Correctness}} indicates problems where the response deviates from objective facts or common knowledge in the field of security. \textbf{\textit{Understandability}} pertains to problems including inappropriate logic, unclear expressions, or incorrect attribution, which make the response difficult to comprehend. \textbf{\textit{Conciseness}} refers to problems related to excessive verbosity or overly complex expressions. \textbf{\textit{Compliance}} involves problems where the response fails to meet the requirements outlined in the task prompt. }

\black{For \black{GPT-4 (ChatGPT)}, most responses contained problems related to \textbf{\textit{Conciseness}}, accounting for 90.18\%. Nearly half of the responses (41.56\%) involved \textbf{\textit{Compliance}}-related problems. Furthermore, 27.08\% of the responses showed problems with \textbf{\textit{Understandability}} and problems related to \textbf{\textit{Correctness}} were the least frequent, occurring in only 14.08\% of the responses. In the responses of DeepSeek-R1, \textbf{\textit{Conciseness}} problems were also the most prevalent, accounting for 84.29\%. However, the frequencies of problems related to \textbf{\textit{Compliance}} and \textbf{\textit{Understandability}} were significantly lower than those of \black{GPT-4 (ChatGPT)}, at 4.51\% and 9.54\%, respectively. Moreover, DeepSeek-R1 was much more prone to \textbf{\textit{Correctness}} problems than \black{GPT-4 (ChatGPT)}, with a rate of 33.66\%.} The typical types of problems within each theme are described in detail below.

\textbf{Correctness:} \black{The responses revealed two types of incorrectness: \textit{Incorrect fact} (\black{GPT-4 (ChatGPT)}: 6.25\%, DeepSeek-R1: 31.81\%) and \textit{Incorrect concept} (\black{GPT-4 (ChatGPT)}: 10.13\%, DeepSeek-R1: 2.39\%).} The responses with \textit{Incorrect fact} include: incorrect line numbers of the code, function names not matching the source code provided, names not corresponding to the CWE-ID, and the described code content not matching the actual scenario. \textit{Incorrect concept} indicates a misunderstanding of the concept of a certain CWE-ID or a specific security defect type. As shown in the example below (Example 1), the inaccurate comparison logic described in the response was mistakenly categorized as CWE-682, thus the response is considered to exhibit the \textit{Incorrect concept} problem. \black{Compared with \black{GPT-4 (ChatGPT)}, DeepSeek-R1 exhibited a lower frequency of \textit{Incorrect concept} problems, but its \textit{Incorrect fact} rate was significantly higher. We observed that during security code reviews, \black{GPT-4 (ChatGPT)} tended to provide the range of code lines where defects were located, while DeepSeek-R1 often identified specific code lines and their corresponding code, making it more prone to errors in details. This may contribute to the higher Incorrect fact rate of DeepSeek-R1 compared to \black{GPT-4 (ChatGPT)}. When applying LLMs to real-world security code reviews, we expect models to provide both specific and accurate factual information. Therefore, how to enhance the capability of LLMs to retain code details with precision requires further exploration.}

\begin{lstlisting}[label={lst:sample1},breaklines=true,breakatwhitespace=true,captionpos=b,basicstyle=\scriptsize\ttfamily,escapeinside=||,frame=single]
|\textbf{Example 1 - Incorrect concept}|
|\textbf{\black{GPT-4 (ChatGPT)}:}| ...**CWE-682 (|\textbf{Incorrect Calculation}|):**
**Issue:** The method `_inventory_has_changed` uses a comparison logic that might not be entirely accurate in all scenarios...
\end{lstlisting}

\textbf{Understandability:} \black{Problems that reduce understandability of responses include \textit{Vague statement} (\black{GPT-4 (ChatGPT)}: 21.38\%, DeepSeek-R1: 3.58\%), \textit{Incoherent expression} (\black{GPT-4 (ChatGPT)}: 1.19\%, DeepSeek-R1: 0.53\%), \textit{Logical contradiction} (\black{GPT-4 (ChatGPT)}: 1.11\%, DeepSeek-R1: 1.13\%), and \textit{Groundless speculation} (\black{GPT-4 (ChatGPT)}: 5.62\%, DeepSeek-R1: 4.11\%).} For \black{GPT-4 (ChatGPT)}, \textit{Vague statement} is the most prevalent one. It refers to a situation where the description of the detected security defect is not specific enough, failing to pinpoint the defect to a specific location or providing an overly general fix suggestion. An example (Example 2) of \textit{Vague statement} is shown below, in which \black{GPT-4 (ChatGPT)} did not provide a concrete location or line number for the identified security defect in the response. While for DeepSeek-R1, the frequency of \textit{Vague statement} was considerably lower than that of \black{GPT-4 (ChatGPT)}. We speculated that its reasoning procedure under the `think' mode somewhat contributed to generating specific and detailed responses.

\begin{lstlisting}[label={lst:sample2},breaklines=true,breakatwhitespace=true,captionpos=b,basicstyle=\scriptsize\ttfamily,escapeinside=||,frame=single]
|\textbf{Example 2 - Vague statement}|
|\textbf{\black{GPT-4 (ChatGPT)}:}| ...
**CWE-703(Improper Check or Handling of Exceptional Conditions)**
**General Concern**: |\textbf{Throughout}| the code, there are instances where error conditions might not be handled comprehensively...
\end{lstlisting}

\textbf{Conciseness:} In all responses, some of them contain \textit{Redundant} content (\black{GPT-4 (ChatGPT)}: 1.14\%, DeepSeek-R1: 39.76\%), some include information \textit{Irrelevant} to the task of security defect detection (\black{GPT-4 (ChatGPT)}: 15.12\%, DeepSeek-R1: 6.03\%), and some contain \textit{Unnecessary} elements (\black{GPT-4 (ChatGPT)}: 86.38\%, DeepSeek-R1: 63.29\%), all of which are problems related to the conciseness of the response. We found that DeepSeek-R1 was more likely than \black{GPT-4 (ChatGPT)} to repeatedly mention the same information in its output, whereas \black{GPT-4 (ChatGPT)} produced content that is unrelated to the target task more frequently. Including \textit{Unnecessary} content, which refers to information that is somewhat related to security code review but does not contribute to detecting security defects, is the most prevalent problem for both \black{GPT-4 (ChatGPT)} and DeepSeek-R1. Such information includes the reasons for the absence of security defects in the current code and general suggestions on security measures and review processes. As illustrated in the example below (Example 3), the response indicates that no defect related to CWE-284 exists in the code and further elaborates on the reason. However, this is not the content that we require \black{GPT-4 (ChatGPT)} to generate in the prompt, hence it is deemed to be \textit{Unnecessary} information.

\begin{lstlisting}[label={lst:sample3},breaklines=true,breakatwhitespace=true,captionpos=b,basicstyle=\scriptsize\ttfamily,escapeinside=||,frame=single]
|\textbf{Example 3 - Unnecessary}|
|\textbf{\black{GPT-4 (ChatGPT)}:}| ...**CWE-284 (Improper Access Control)**: 
- There |\textbf{doesn't}| seem to be any direct issue related to improper access control in the given code...
\end{lstlisting}

\textbf{Compliance:} Four types of non-compliance were identified regarding whether the responses satisfied the requirements in the prompts. Among them, \textit{Detection failure} cases were observed in very few responses (\black{GPT-4 (ChatGPT)}: 0.15\%, DeepSeek-R1: 0.13\%), where LLM explicitly claimed that it could not perform a security check in its response or was stuck in an infinite output loop and failed to complete the security defect detection task. \textit{Task misunderstanding}, which is similarly rare (\black{GPT-4 (ChatGPT)}: 0.00\%, DeepSeek-R1: 0.66\%), indicates that LLM misinterprets the task details in the prompt. For instance, when using DeepSeek-R1 with \(\textbf{P}_{cot-guardrail}\), if the commit message mentioned security defects addressed by the current commit, LLM may mistakenly interpret the task of security code review as explaining the security defects mentioned in the commit message. \textit{Non-compliant output} was present in 20.43\% of \black{GPT-4 (ChatGPT)} responses, in which \black{GPT-4 (ChatGPT)} fails to include the fixed phrase `\textit{No security defects are detected in the code}', when no defects are found. \textit{Incomplete answer} is the most common problem (20.98\%) under this theme for \black{GPT-4 (ChatGPT)}, referring to cases in which the responses lack the line numbers or fix suggestions of detected security defects requested in the prompt. Although \black{GPT-4 (ChatGPT)} showed relatively high frequencies for these two problems, DeepSeek-R1 exhibited much lower rates for them, at 1.86\% and 1.99\%, respectively. Given this, we consider DeepSeek-R1 to exhibit stronger instruction-following capabilities than \black{GPT-4 (ChatGPT)} in security code review. The example below (Example 4) shows that while the potential memory management issues caused by temporary files and directories are described and a fix suggestion is provided, no specific line numbers are given. Hence, it is categorized as an \textit{Incomplete answer}. 

\begin{lstlisting}[label={lst:sample4},breaklines=true,breakatwhitespace=true,captionpos=b,basicstyle=\scriptsize\ttfamily,escapeinside=||,frame=single]
|\textbf{Example 4 - Incomplete answer}|
|\textbf{\black{GPT-4 (ChatGPT)}:}| ### 3. CWE-664
**Potential Issue:** The creation of temporary files and directories (e.g., in function `createQmlrcFile`) ... can potentially lead to resource management issues.
**Solution:** Implement thorough checks and error handling...
\end{lstlisting}

While \cite{kabir2023answers} also evaluated the quality problems in LLM-generated responses, the application domain of LLMs and the distribution of identified problems in their work differ from ours. Specifically, \cite{kabir2023answers} requested ChatGPT to answer programming questions on Stack Overflow \black{and found that ChatGPT makes fewer factual errors compared to conceptual errors. This aligns with our findings for \black{GPT-4 (ChatGPT)}. However, unlike their study, which identified correctness problems in over half (52\%) of the responses, our study applied \black{GPT-4 (ChatGPT)} to security code review, revealing a much lower incidence of correctness issues at 14.81\%. When applying DeepSeek-R1, the frequency of correctness problems increased to 33.66\%. This suggests that the distribution of quality problems in LLM responses may vary significantly across different application domains and model architectures, necessitating distinct optimization directions and strategies in practical application.}

\subsubsection{RQ2.2}
\textcolor{black}{We selected all items related to hallucination from the quality problem types in RQ1.1 and mapped them to the predefined set of hallucination subtypes. The details of mapping, corresponding explanations, and proportion results are shown in Table~\ref{tbl:hallucination}. Since a single response may exhibit multiple types of hallucination simultaneously, the sum of the proportions for each subtype exceeds the proportion of responses containing at least one hallucination. 
\black{Overall, 20.19\% of \black{GPT-4 (ChatGPT)} responses contained hallucinations, whereas DeepSeek-R1 exhibited a higher rate of 36.45\%. The former encountered more \textit{Factuality Hallucination}, while the latter was predominantly affected by \textit{Faithfulness Hallucination}. Specifically, for \black{GPT-4 (ChatGPT)}, \textit{Factual Contradiction} was the most frequent subtype, accounting for 10.13\%, followed by \textit{Context Inconsistency} (6.25\%) and \textit{Factual Fabrication} (5.62\%). In contrast, for DeepSeek-R1, \textit{Context Inconsistency} was the most prevalent subtype (31.81\%), with \textit{Factual Fabrication} and \textit{Factual Contradiction} following at 4.11\% and 2.39\%, respectively.} To alleviate these hallucinations, Retrieval-Augmented Generation (RAG) can be employed to incorporate external factual information~\citep{tonmoy2024comprehensive, lewis2020retrieval}, and post-training can also be applied to enhance the model's capability to capture and retain contextual knowledge~\citep{kumar2025llm}.}

\begin{table*}
\scriptsize
\begin{threeparttable}
\caption{\textcolor{black}{Mapping of hallucination and quality problem types with corresponding proportions}}
 \label{tbl:hallucination}
\begin{tabular}{M{1.8cm}|M{2.8cm}|M{2.5cm}|p{5.8cm}|p{0.9cm}|p{0.9cm}}
\hline
\multirow{2}{*}{\textbf{Halluc. Type*}}    & \multirow{2}{*}{\textbf{Subtype}}  & \multirow{2}{*}{\textbf{QP Type*}}   &\multirow{2}{*}{\textbf{Explanation}}   &\multicolumn{2}{c}{\textbf{Prop.*}}\\\cline{5-6}
~ & & & & GPT-4$^\dagger$ & DS-R1$^\dagger$ \\\hline
\multirow{2}{*}{\makecell[cl]{Factuality\\Hallucination}} & Factual Contradiction  &Incorrect Concept  &In \textit{Incorrect Concept} cases, the model's output contradicts the authoritative definition of the corresponding CWE, thus categorized as \textit{Factual Contradiction}. &\textbf{10.13\%} &2.39\%\\\cline{2-6}
~   &Factual Fabrication  &Groundless Speculation&\textit{Groundless speculation} refers to subjective assumptions made without contextual evidence or knowledge support. As it involves fabricated factual content, it is categorized as \textit{Factual Fabrication}. &\textbf{5.62\%} &4.11\%\\\hline
\multirow{3}{*}{\makecell[l]{Faithfulness\\Hallucination}} &Instruction Inconsistency  &Detection Failure, Task Misunderstanding  &\textit{Detection Failure} refers to cases where the model fails to perform the security checks or misinterprets the task, leading to deviations from the instruction. Therefore, it is categorized as \textit{Instruction Inconsistency}.&0.15\% &\textbf{0.80\%}\\\cline{2-6}
~   &Context Inconsistency  &Incorrect Fact  & \textit{Incorrect Fact} denotes cases where the line numbers, identifiers, or code content generated by the model do not align with the provided code context, thus categorized as \textit{Context Inconsistency}. &6.25\% &\textbf{31.81\%}\\\cline{2-6}
~   &Logical Inconsistency  &Logical Contradiction &\textit{Logical Contradiction} refers to cases where the model's reasoning about a security defect is internally contradictory, thus classified as \textit{Logical Inconsistency}. &1.11\% &1.13\%\\\hline
\multicolumn{3}{l|}{\textbf{Any Hallucination}}  & & 20.19\% &\textbf{36.45\%}\\\hline
\end{tabular}
\begin{tablenotes}
    \item[\textbf{*}] Abbreviations: Halluc. = Hallucination, QP = Quality Problem, Prop. = Proportion.
    \item[$^\dagger$] GPT-4 = GPT-4 (ChatGPT), DS-R1 = DeepSeek-R1
\end{tablenotes}
\end{threeparttable}
\end{table*}
\subsection{Factors Influencing LLMs (RQ3)}
\label{sec: rq3_result}
\subsubsection{Model Stability}
To assess the adequacy of the fit of regression models for \textit{\black{GPT-4 (ChatGPT)}+\(\textbf{P}_{cid}\)} and \textit{DeepSeek-R1+\(\textbf{P}_{cot-guardrail}\)} \black{(hereafter referred to as \textit{\black{GPT-4 (ChatGPT)}} and \textit{DeepSeek-R1})}, we utilized Nagelkerke's Pseudo $R^2$~\citep{nagelkerke1991note}. This yielded values of \black{0.0923} for \black{GPT-4 (ChatGPT) and 0.1145} for DeepSeek-R1, which are relatively lower than the results of regression analysis in other fields reported in e.g.,~\citep{paul2021security, turzo2022towards}. Nonetheless, through a log-likelihood ratio test using the \textit{anova} function of the \textit{ordinal} R package~\citep{christensen2015package}, we found that our regression models exhibit significant explanatory power over a null model: \black{for \black{GPT-4 (ChatGPT)}, $LR\,chi2=\black{107.47}$, $P<0.0001$; for DeepSeek-R1: $LR\,chi2=\black{156.37}$, $P<0.0001$}. Given that our goal is to construct an inferential regression model rather than a predictive one, these two regression models with relatively lower $R^2$ can still provide valid insights into the relationship between explanatory variables and response variables~\citep{allison2014prediction}.

\begin{table*}
\small
\begin{threeparttable}
\caption{Explanatory powers of factors to the performance of GPT-4 (ChatGPT) under \({P}_{cid}\) in security code review}
 \label{tbl:explanatory_gs}
\begin{tabular}{|c|l|c|c|c|c|c|c|c|c|}
\hline
\multicolumn{2}{|l|}{\multirow{2}{*}{\textbf{Factor (:Reference)}}}  &\multirow{2}{*}{\textbf{OR (95\% CI)}} & \multirow{2}{*}{\textbf{D.F.}} &\multirow{2}{*}{\textbf{$\chi^2$}} &\multirow{2}{*}{\textbf{Pr(>Chisq)$^{\dagger}$}}  &\multicolumn{4}{c|}{\textbf{AME}} \\\cline{7-10}
\multicolumn{2}{|l|}{} & & &  & &M &U &H &I \\\hline 
\multicolumn{2}{|l|}{TK}                                                     & -                 &2                  &40.11          &$<.0001^{***}$   &0.1002 &-0.0518 &-0.0265 &-0.0219  \\\hline
\multicolumn{2}{|l|}{PT}                                                  &0.98 (0.87, 1.10)   &1                  &0.43           &0.9197  &0.0054 &-0.0028 &-0.0014 &-0.0012  \\\hline
\multicolumn{1}{|l|}{CMT}  &Qt:Openstack                                  &1.42 (1.07, 1.89)   &1                  &4.69           &0.2122  &-0.0790 &0.0415 &0.0206 &0.0169  \\\hline
\multicolumn{1}{|l|}{FT}   &Source:Auxiliary                              &\textcolor{green}{2.65} (1.69, 4.18)   &1                  &17.89           & 0.0002***  &-0.2231 &0.1432 &0.0456 &0.0344  \\\hline
\multicolumn{1}{|l|}{\multirow{2}{*}{ASR}} &-1:0        &\textcolor{green}{1.09} (0.80, 1.40)   &\multirow{2}{*}{2} &\multirow{2}{*}{16.89} &\multirow{2}{*}{0.0019**}  &-0.0129 &0.0069 &0.0033 &0.0027 \\\cline{2-3}\cline{7-10}
~                                                                   &1:0        &\textcolor{green}{2.93} (1.73, 4.97)   &~                  &~                       &~   &-0.2129 &0.0547 &0.0789 &0.0793\\\hline
\multicolumn{2}{|l|}{AE}                                          &0.94 (0.83, 1.05)   & 1                 &1.46           &0.7154      &0.0149 &-0.0077 &-0.0039 &-0.0033 \\\hline
\multicolumn{2}{|l|}{CT}                                                    &1.14 (1.01, 1.28)   & 1                 &4.59           &0.2122  &-0.0291 &0.0150 &0.0077 &0.0064\\\hline
\multicolumn{1}{|l|}{\multirow{4}{*}{SDT}}   &Crash:Thread       &0.98 (0.73, 1.30)   &\multirow{4}{*}{4} &\multirow{4}{*}{13.79} &\multirow{4}{*}{0.0641}  &0.0057 &-0.0029 &-0.0015 &-0.0012  \\\cline{2-3}\cline{7-10}
~                                                           &Memory:Thread      &0.80 (0.55, 1.17)   &~      &~     &~   &0.0499 &-0.0278 &-0.0123 &-0.0098  \\\cline{2-3}\cline{7-10}
~                                                           &Permission:Thread  &0.80 (0.52, 1.22)  &~      &~     &~   &0.0508 &-0.0283 &-0.0126 &-0.0100  \\\cline{2-3}\cline{7-10}
~                                                           &Resource:Thread    &1.52 (0.99, 2.33)   &~      &~     &~   &-0.0899 &0.0372 &0.0279 &0.0247  \\\hline
\multicolumn{2}{|l|}{CPT}                                                &1.10 (0.99, 1.23)   &1                  &3.96  &0.2336  &-0.0218  &0.0113  &0.0057 &0.0048 \\\hline
\multicolumn{2}{|l|}{AR}                                           &0.92 (0.82, 1.04)   &1                  &0.55   &0.9197     &0.0178  &-0.0092   &-0.0047 &-0.0039  \\\hline
\multicolumn{1}{|l|}{AHC} &1:0                                         &0.87 (0.61, 1.24)   &1                  &1.81   &0.7154  &0.0315 &-0.0170 &-0.0080 &-0.0065  \\\hline
\end{tabular}
\begin{tablenotes}
    \item[$^{\dagger}$] Statistical significance based on the Wald $\chi^2$ test (\textit{p-values} adjusted using the Holm method~\citep{holm1979simple}): \noindent \\$^* p <0.05$; $^{**} p <0.01$; $^{***} p <0.001$
\end{tablenotes}
\end{threeparttable}
\end{table*}
\begin{table*}
\small
\begin{threeparttable}
\caption{Explanatory powers of factors to the performance of DeepSeek-R1 under \({P}_{cot-guardrail}\) in security code review}
 \label{tbl:explanatory_ds}
\begin{tabular}{|c|l|c|c|c|c|c|c|c|c|}
\hline
\multicolumn{2}{|l|}{\multirow{2}{*}{\textbf{Factor (:Reference)}}}  &\multirow{2}{*}{\textbf{OR (95\% CI)}} & \multirow{2}{*}{\textbf{D.F.}} &\multirow{2}{*}{\textbf{$\chi^2$}} &\multirow{2}{*}{\textbf{Pr(>Chisq)$^{\dagger}$}}  &\multicolumn{4}{c|}{\textbf{AME}} \\\cline{7-10}
\multicolumn{2}{|l|}{} & & &  & &M &U &H &I \\\hline 
\multicolumn{2}{|l|}{TK}                                                     & -                 &2                  &30.31          &$<.0001^{***}$   &0.0933 &-0.0340 &-0.0113 &-0.0479  \\\hline
\multicolumn{2}{|l|}{PT}                                                  &1.13 (0.97, 1.33)   &1                  &0.56           &1.0000  &-0.0079 &0.0075 &0.0002 &0.0002  \\\hline
\multicolumn{1}{|l|}{CMT}  &Qt:Openstack                                  &\textcolor{green}{1.45} (1.10, 1.91)   &1                  &8.85           &0.0293*  &-0.0746 &0.0371 &0.0074 &0.0299  \\\hline
\multicolumn{1}{|l|}{FT}   &Source:Auxiliary                              &0.66 (0.41, 1.08)   &1                  &5.43           & 0.1384  &0.0762 &-0.0290 &-0.0089 &-0.0382  \\\hline
\multicolumn{1}{|l|}{\multirow{2}{*}{ASR}} &-1:0        &\textcolor{green}{1.31} (1.01, 1.69)   &\multirow{2}{*}{2} &\multirow{2}{*}{12.41} &\multirow{2}{*}{0.0266*}  &-0.0527 &0.0250 &0.0055 &0.0222 \\\cline{2-3}\cline{7-10}
~                                                                   &1:0        &\textcolor{green}{2.78} (1.62, 4.77)   &~                  &~                       &~   &-0.1721 &0.0362 &0.0235 &0.1124\\\hline
\multicolumn{2}{|l|}{AE}                                          &0.97 (0.87, 1.09)   & 1                 &0.54           &1.0000      &0.0051 &-0.0025 &-0.0005 &-0.0021 \\\hline
\multicolumn{2}{|l|}{CT}                                                    &1.00 (0.90, 1.11)   & 1                 &0.04           &1.0000  &-0.0007 &0.0003 &0.0001 &0.0003\\\hline
\multicolumn{1}{|l|}{\multirow{4}{*}{SDT}}   &Crash:Thread       &1.46 (0.77, 2.78)   &\multirow{4}{*}{4} &\multirow{4}{*}{41.38} &\multirow{4}{*}{$<.0001^{***}$}  &-0.0217 &0.0112 &0.0019 &0.0089  \\\cline{2-3}\cline{7-10}
~                                                           &Memory:Thread      &1.37 (0.53, 3.58)   &~      &~     &~   &0.0133 &-0.0310 &0.0034 &0.0142  \\\cline{2-3}\cline{7-10}
~                                                           &Permission:Thread  &3.65 (1.20, 11.13)  &~      &~     &~   &-0.1462 &0.0324 &0.0189 &0.0949  \\\cline{2-3}\cline{7-10}
~                                                           &Resource:Thread    &0.52 (0.22, 1.20)   &~      &~     &~   &-0.0304 &0.0073 &0.0042 &0.0190  \\\hline
\multicolumn{2}{|l|}{CPT}                                                &1.28 (1.00, 1.64)   &1                  &13.96  &0.0022**  &-0.0545  &0.0261  &0.0055 &0.0229 \\\hline
\multicolumn{2}{|l|}{AR}                                           &1.00 (0.89, 1.12)   &1                  &0.15   &1.0000     &0.0007  &-0.0003   &-0.0001 &-0.0003  \\\hline
\multicolumn{1}{|l|}{AHC} &1:0                                         &1.50 (1.04, 2.15)   &1                  &6.72   &0.0764  &-0.0758 &0.0297 &0.0087 &0.0374  \\\hline
\multicolumn{2}{|l|}{SDT × TK}                                & -                  &8                  &15.99  &0.2554  &\multicolumn{4}{c|}{-}  \\\hline
\multicolumn{2}{|l|}{SDT × PT}                             & -                  &4                  &9.49   &0.2554  &\multicolumn{4}{c|}{-} \\\hline
\multicolumn{2}{|l|}{SDT × CPT}                           & -                  &4                  &15.83  &0.0293*  &\multicolumn{4}{c|}{-}  \\\hline
\end{tabular}
\begin{tablenotes}
    \item[$^{\dagger}$] Statistical significance based on the Wald $\chi^2$ test (\textit{p-values} adjusted using the Holm method~\citep{holm1979simple}): \noindent \\$^* p <0.05$; $^{**} p <0.01$; $^{***} p <0.001$
\end{tablenotes}
\end{threeparttable}
\end{table*}
\begin{table}
\small
\begin{threeparttable}
\caption{Statistical power estimates for different effect sizes}
 \label{tbl:power}
 \setlength{\tabcolsep}{5pt}
\begin{tabular}{|c|c|c|c|c|c|c|}
\hline
\multicolumn{1}{|l|}{\multirow{3}{*}{\textbf{Factor}}}  &\multicolumn{6}{c|}{\textbf{Cohen's d}}  \\\cline{2-7}
\multicolumn{1}{|l|}{}  &\multicolumn{3}{c|}{GPT-4 (ChatGPT)} &\multicolumn{3}{c|}{DeepSeek-R1} \\\cline{2-7}
\multicolumn{1}{|l|}{}  &0.2 &0.5 &0.8 &0.2 &0.5 &0.8 \\\hline 
\multicolumn{1}{|l|}{TK}                                                   &$>.99$ &$>.99$ &$>.99$ &.15 &.35&.45  \\\hline
\multicolumn{1}{|l|}{PT}                                                  &$>.99$ &$>.99$ &$>.99$ &.08 &.20&.28  \\\hline
\multicolumn{1}{|l|}{CMT}                                                &.87 &$>.99$ &$>.99$ &.66 &.96&.98  \\\hline
\multicolumn{1}{|l|}{FT}                                                &.42 &.95 &.99 &.27 &.65&.76  \\\hline
\multicolumn{1}{|l|}{ASR}                               &$>.99$ &$>.99$ &$>.99$ &$>.99$ &$>.99$&$>.99$ \\\hline
\multicolumn{1}{|l|}{AE}                                           &$>.99$ &$>.99$ &$>.99$ &$>.99$&$>.99$&$>.99$ \\\hline
\multicolumn{1}{|l|}{CT}                                                     &$>.99$ &$>.99$ &$>.99$ &$>.99$&$>.99$&$>.99$ \\\hline
\multicolumn{1}{|l|}{SDT}                                &$>.99$ &$>.99$ &$>.99$ &$>.99$&$>.99$&$>.99$  \\\hline
\multicolumn{1}{|l|}{CPT}                                                 &$>.99$ &$>.99$ &$>.99$ &.08 &.21 &.28 \\\hline
\multicolumn{1}{|l|}{AR}                                           &$>.99$ &$>.99$ &$>.99$ &$>.99$&$>.99$&$>.99$ \\\hline
\multicolumn{1}{|l|}{AHC}                                     &.52 &.96  &.99 &.34&.73&.82 \\\hline
\multicolumn{1}{|l|}{SDT × TK}                                &\multicolumn{3}{c|}{-} &$>.99$&$>.99$&$>.99$ \\\hline
\multicolumn{1}{|l|}{SDT × PT}                              &\multicolumn{3}{c|}{-} &$>.99$&$>.99$&$>.99$ \\\hline
\multicolumn{1}{|l|}{SDT × CPT}                           &\multicolumn{3}{c|}{-} &$>.99$&$>.99$&$>.99$ \\\hline
\end{tabular}
\end{threeparttable}
\end{table}

\begin{figure*}
    \centering
    \includegraphics[width=.9\linewidth]{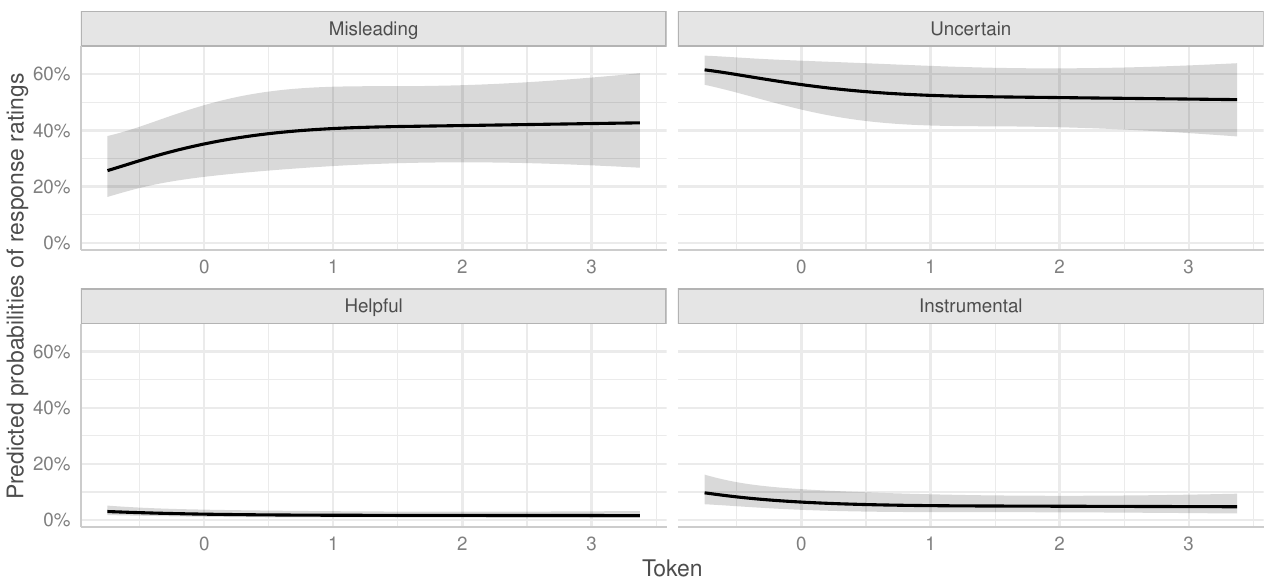}
    \caption{Predicted probabilities for the response variable by `TK' in the regression model of DeepSeek-R1.}   
    \label{fig:tk_or_ds}
\end{figure*}

\begin{figure*}
    \centering
    \includegraphics[width=.9\linewidth]{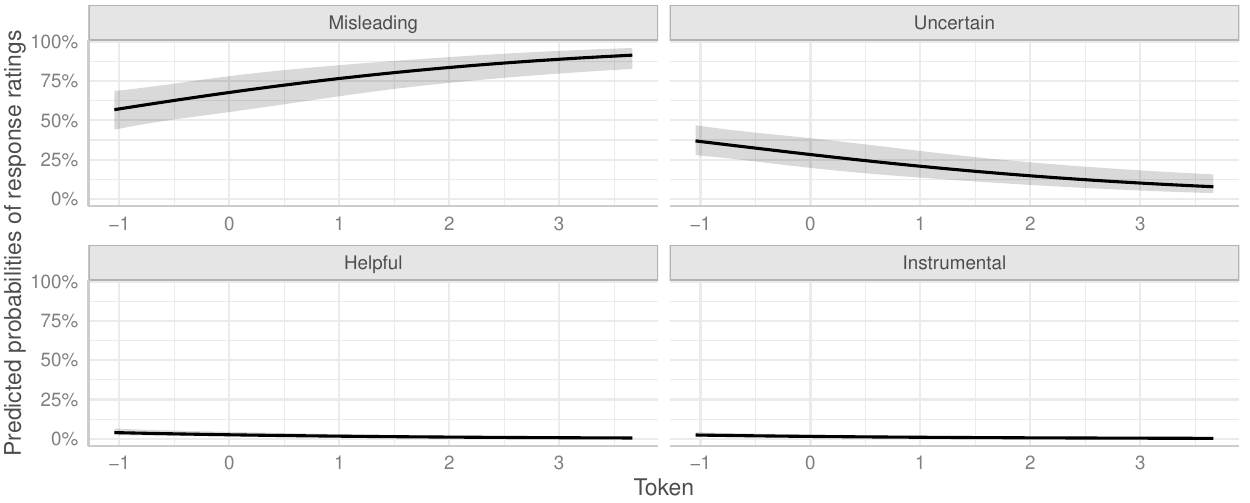}
    \caption{Predicted probabilities for the response variable by `TK' in the regression model of GPT-4 (ChatGPT).}   
    \label{fig:tk_or_gs}
\end{figure*}

\begin{figure*}
    \centering
    \includegraphics[width=.9\linewidth]{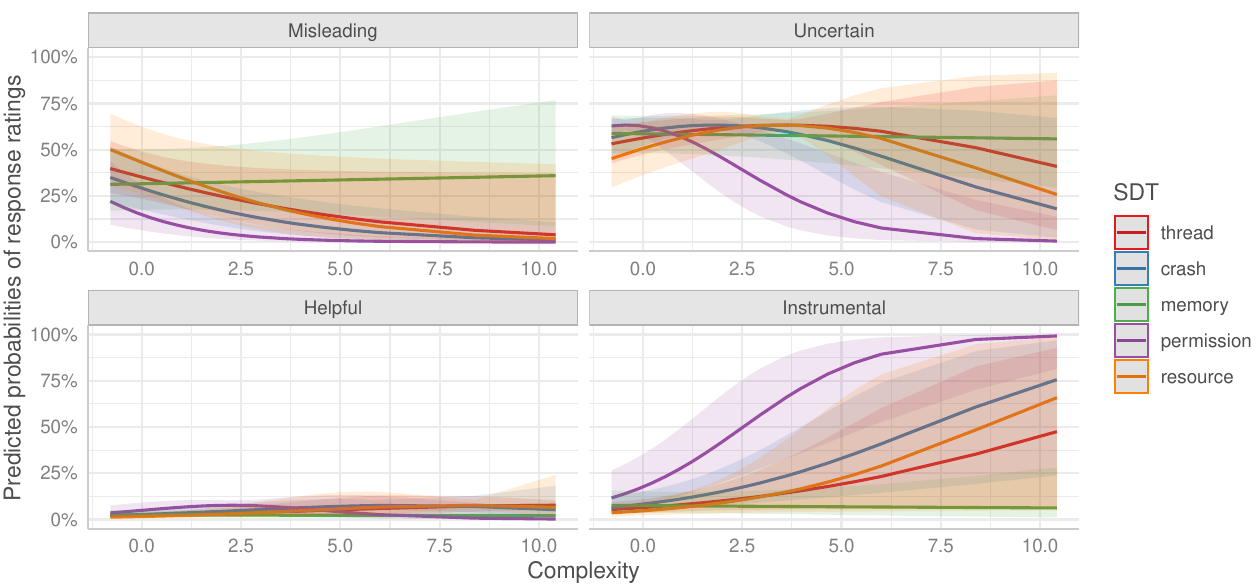}
    \caption{Predicted probabilities for the response variable by `CPT' conditional on `SDT' in the regression model of DeepSeek-R1.}   
    \label{fig:cpt_inter_ds}
\end{figure*}

\subsubsection{Explanatory Power of Each Factor}
The Wald statistic is employed to evaluate the explanatory power exerted by each factor on the response variable. A larger Wald $\chi^2$ with a smaller \textit{p-value} indicates that the variable has more significant predictive power for the regression model. We calculated the $\chi^2$ and \textit{p-value} of each factor \black{(see Table~\ref{tbl:explanatory_gs} and Table~\ref{tbl:explanatory_ds})} and adopted the commonly used significance level of 0.05 to evaluate their impact. \black{Considering the multiplicity in multiple regression~\citep{anderson2023multiplicity}, \textit{p-values} for the explanatory variables were adjusted using the Holm stepwise procedure~\citep{holm1979simple} for multiple tests correction.}

\black{As shown in Table~\ref{tbl:explanatory_gs}, \textbf{for GPT-4 (ChatGPT)}, `Token' exhibited the highest explanatory power on the fitted model, with the highest $\chi^2$ (\black{40.11}) and the lowest \textit{p-value} ($<0.0001$). This was followed by `FileType' and `AnnotationisSecurityRelated', both of which exhibited significant contributions, with $\chi^2=\black{17.89}$ ($p=\black{0.0002}$) and $\chi^2=16.89$ ($p=0.0019$), respectively. \textbf{For DeepSeek-R1}, we can see from Table~\ref{tbl:explanatory_ds} that among the three interaction terms, only the `SecurityDefectType × Complexity' term exhibited significant explanatory power, with $\chi^2=15.83$ and $p=0.0293$, suggesting that the effect of `Complexity' may vary across different types of security defects. The corresponding main effects of `SecurityDefectType' and `Complexity' yielded $\chi^2=41.83$ ($p<0.0001$) and $\chi^2 = 13.96$ ($p=0.0022$), respectively. Among the other predictors, `Token' demonstrated the strongest explanatory power, with $\chi^2=30.31$ and $p<0.0001$. `Community' and `AnnotationisSecurityRelated' also exhibited statistical significance, with $\chi^2 = 8.85$ ($p=0.0293$) and $\chi^2 = 12.41$ ($p=0.0266$).}



\black{Factors not discussed above have \textit{p-values} exceeding the 0.05 threshold, indicating that no statistically significant associations between these factors and LLM performance in security defect detection were observed. To assess the reliability of these non-significant findings, we further analyzed them with the power analysis results reported in Table~\ref{tbl:power}. \textbf{For GPT-4 (ChatGPT)}, `Position', `AuthorExperience', `Commit', `SecurityDefectType', `Complexity', and `AnnotationRatio' all achieved statistical power above 0.99 under small effect assumptions, suggesting that the sample size ($N=1263$) is sufficient to defect small effects of these factors. In addition, `AnnotationHasCode' attained a power of 0.96 for a medium effect size, indicating adequate sensitivity to detect such effects. \textbf{For DeepSeek-R1}, `AuthorExperience', `Commit', `AnnotationRatio', as well as the interaction terms `SecurityDefectType × Token' and `SecurityDefectType × Position' also exhibited power exceeding 0.99 for small effect sizes. Meanwhile, `Community' achieved a power of 0.96 under medium effect, and `AnnotationHasCode' reached 0.82 at the large effect level. These results suggest that the non-significance observed in these factors may reflect a true absence of effect or imply that their effects are smaller than their corresponding thresholds mentioned above. However, for DeepSeek-R1, `Position' and `FileType' exhibited power below 0.80 across all assumed effect sizes, suggesting that their non-significant results may be attributed to insufficient sample size and need further validation with more data. Notably, for DeepSeek-R1, while `Token' and `Complexity' showed statistical significance, their power remained below 0.80 across all effect levels, raising concerns about potential false positives and limited robustness of the observed significance.}

\begin{table*}

\begin{threeparttable}
\caption{The odds ratios and \textit{p-values} of the interaction terms for DeepSeek-R1}
 \label{tbl:interaction}
\begin{tabular}{|c|c|c|c|c|c|c|}
\hline
\multirow{2}{*}{SDT}  &\multicolumn{2}{c|}{TK} &\multicolumn{2}{c|}{PT} &\multicolumn{2}{c|}{CPT} \\\cline{2-7}
~  &OR (95\% CI) &Pr(>|z|) &OR (95\% CI) &Pr(>|z|) &OR (95\% CI) &Pr(>|z|) \\\hline
Thread      &\textcolor{red}{0.66} (0.47, 0.92)    &0.0389*    &1.13 (0.97, 1.33)  &0.5020     &1.28 (1.00, 1.64)   &0.1233\\\hline
Crash       &0.86 (0.56, 1.32)    &0.9705     &1.07 (0.86, 1.33)  &0.9360     &\textcolor{green}{1.40} (1.13, 1.74)   &0.0078**\\\hline
Memory      &\textcolor{red}{0.39} (0.19, 0.77)    &0.0276*    &0.73 (0.52, 1.02)  &0.3210     &0.98 (0.84, 1.15)   &0.8179\\\hline
Permission  &0.91 (0.39, 2.09)    &0.9705     &1.22 (0.83, 1.80)  &0.9360     &\textcolor{green}{1.84} (1.33, 2.55)   &0.0010**\\\hline
Resource    &\textcolor{red}{0.35} (0.19, 0.65)    &0.0041**   &0.87 (0.64, 1.19)  &0.9360     &1.42 (1.01, 1.99)   &0.1233\\\hline
\end{tabular}
\begin{tablenotes}
    \item[$^{\dagger}$] Statistical significance based on the Wald $\chi^2$ test (\textit{p-values} adjusted using the Holm method~\citep{holm1979simple}): \noindent \\$^* p <0.05$; $^{**} p <0.01$; $^{***} p <0.001$
\end{tablenotes}
\end{threeparttable}
\vspace{-1em}
\end{table*}

\subsubsection{Relationships between Factors and Responses}
The odds ratio (OR) is commonly used to measure the relationships between explanatory and response variables~\citep{paul2021security, thongtanunam2017review}. In this study, the OR represents the odds of LLM generating a better response in security code review versus not generating a better response. For categorical explanatory variables, the OR compares each category to a reference category. For interval variables, the OR reflects the effect of a one-unit increase in the explanatory variable. \black{We utilized the \textit{summary} function of the \texttt{rms} package~\citep{harrell2017package} and the \textit{confint} function to calculate the OR with its 95\% confidence interval for each factor. Additionally, average marginal effects (AME) were computed using the \textit{avg\_slopes} function of the \texttt{marginaleffects} package to facilitate a more granular interpretation of the predictors' influence. These results are all presented in Table~\ref{tbl:explanatory_gs} and Table~\ref{tbl:explanatory_ds}. Considering that the complex effects of non-linear and interaction terms cannot be displayed by a single odds ratio, we visualized their effects using predicted probability curves of the response variable (see Fig.~\ref{fig:tk_or_ds}, Fig.~\ref{fig:tk_or_gs} and Fig.~\ref{fig:cpt_inter_ds}), thus providing a more intuitive view of how these factors influence LLM performance. To further investigate the interaction effects, we used the \textit{emtrends} function from the \texttt{emmeans} package to calculate the simple slopes of `Token', `Position', and `Complexity' at different levels of security defect types. The detailed results are presented in Table~\ref{tbl:interaction}.}

\black{As presented in Fig.~\ref{fig:tk_or_ds} and Fig.~\ref{fig:tk_or_gs}, \textbf{for both DeepSeek-R1 and GPT-4 (ChatGPT)}, as the `token` factor increases, the probability of the response variable being predicted as the \textit{Instrumental}, \textit{Helpful}, and \textit{Uncertain} categories all show a decreasing trend, while the predicted probability of \textit{Misleading} obviously increases. The results indicate that in code files with more tokens, LLMs tend to provide worse security defect detection results. However, for DeepSeek-R1, given the low statistical power of the `Token' variable, although the relationship between `Token' and the response variable is similar to that observed for GPT-4 (ChatGPT), additional data is required for further validation. As presented in Table ~\ref{tbl:explanatory_gs} and Table ~\ref{tbl:explanatory_ds}, for both DeepSeek-R1 and GPT-4 (ChatGPT), the ORs of `AnnotationisSecurityRelated', together with their 95\% CIs, are generally greater than 1, suggesting that both LLMs are more likely to detect security defects in code files containing security-related annotations. Moreover, \textbf{for GPT-4 (ChatGPT)}, the `FileType' factor has an OR of 2.65 (95\% CI: from 1.69 to 4.18), indicating that GPT-4 (ChatGPT) has a higher likelihood of detecting security defects in source code with functional logic. \textbf{For DeepSeek-R1}, the OR value for `Community' is 1.45 (95\% CI: from 1.10 to 1.91), suggesting better LLM performance on code files originate from the Qt community, rather than from OpenStack. For the variables discussed above, the AME results are consistent with the corresponding OR estimates and predicted probability curves. Specifically, for variables with positive effects ($OR > 1$), as their values increase (or compared with the reference category for categorical variables), the probability of the response being classified as ``Misleading'' tends to decrease. At the same time, the probability of higher response categories (i.e., ``Uncertain'', ``Helpful'' and ``Instrumental'') increases. Conversely, variables with negative effects ($OR < 1$) exhibit the opposite trend. These findings further support the reliability of the identified directions of influence for these factors on LLM performance in security defect detection.} 

\black{Regarding the interaction term `SecurityDefectType × Complexity' for DeepSeek-R1, as shown in Fig.~\ref{fig:cpt_inter_ds}, when the `SecurityDefectType' is ``Memory'', the predicted probabilities of each response category have no noticeable changes with the increase of `Complexity'. While when the `SecurityDefectType' takes values other than ``Memory'', higher complexity is associated with a decrease in the probabilities of the \textit{Misleading} and \textit{Uncertain} categories and an increase in the \textit{Instrumental} category. This suggests that DeepSeek-R1 demonstrates stronger capabilities in security defect detection for files with more code complexity. Table~\ref{tbl:interaction} presents the results consistent with the above analysis. Specifically, for memory-related security defects, the OR of `Complexity' is 0.98 (95\% CI: from 0.84 to 1.15), showing no clear trend or significant impact. For all other security defect types, the ORs together with their 95\% CIs are all greater than 1, indicating a positive association between code complexity and the performance of DeepSeek-R1 in security defect detection. This result contrasts with the consensus in previous work that as complexity of software grows, security defects are becoming more challenging to detect~\citep{sheng2024lprotector}. A possible explanation is that we utilized McCabe's Cyclomatic Complexity~\citep{mccabe1976complexity} to measure the `Complexity' factor, which is not always positively correlated with the actual difficulty of detection. Code with higher cyclomatic complexity may have richer control-flow structures and retain more contextual information within the single code file, which could better support DeepSeek-R1's reasoning procedure during its think mode. However, memory-related security defects, such as \textit{Integer Overflow} and \textit{Buffer Overflow}, are typically characterized by local patterns that can be identified without relying on global structures, thus appearing to be less affected by the `Complexity' factor.}

\black{Consistent with prior work~\citep{Sovrano_2025}, we observe that as the number of tokens increases, the detection performance of LLMs tend to decline. However, our findings regarding the `Position' factor differ from those reported in previous studies. Existing research suggest that LLMs tend to significantly underperform when detecting security defects located toward the end of code files~\citep{Sovrano_2025, rafi2024order}. However, in our analysis, the effect of `Position' is not statistically significant with the 95\% confidence intervals of the odds ratios including 1 for both DeepSeek-R1 and GPT-4 (ChatGPT) (see Table~\ref{tbl:interaction} and Table~\ref{tbl:explanatory_gs}). For DeepSeek-R1, the `Position' variable exhibited low statistical power across all tested effect sizes, indicating that the sample size ($N=1263$) may be insufficient to reliably detect the effect of `Position'. This limited power may partially explain the non-significance of `Position'. While for GPT-4 (ChatGPT), the sample size ($N=1509$) is sufficient to detect even small effects of `Position', suggesting that insufficient statistical power may not be the primary cause of the non-significance of `Position'. A possible explanation lies in differences in the distribution of security defect types between this study and prior work. Although no statistically significant interaction between `SecurityDefectType' and `Position' was observed, this does not preclude the existence of such effects in defect types that are underrepresented or not included in this study, such as \textit{XSS}, \textit{SQL Injection}, and \textit{Path Traversal} (as discussed in ~\cite{Sovrano_2025}). Furthermore, we grouped security defects into five broad categories in the data used for model fitting, which may mask fine-grained variations in the effect of `Position' across specific defect types. Therefore, future work will focus on collecting additional data or conducting controlled perturbation tests, to better understand the source of these discrepancies and to validate the comprehensive effect of the `Position' factor on LLM performance in security defect detection.}

\section{Discussion} 
\label{sec:implications}

\textbf{\textit{A multi-layered review strategy is suggested to apply LLMs to security code review.}} As the results of RQ1 indicate, the existing general-purpose LLMs still fall significantly short of reaching the effectiveness of manual security code review. However, LLMs can serve as auxiliary tools to assist reviewers. We recommend a multi-layered review strategy to fully leverage the capability of LLMs. The findings of RQ3 suggest that LLMs are more adept at identifying security defects in code file with fewer tokens and higher cyclomatic complexity. Utilizing these two factors, a lightweight model can be trained to classify the code file to be reviewed into the low-difficulty or high-difficulty groups. Firstly, LLMs are utilized to conduct an initial automatic review of these files. Then, for low-difficulty code, reviewers quickly confirm the security defects identified by LLMs. Code in which LLMs did not detect any security defects would then be resampled and rechecked to confirm its security. For high-difficulty code, reviewers utilized the detection results of LLMs as a reference, combined with their own domain knowledge, to perform a more thorough and in-depth analysis of the submitted code. This approach aims to improve the efficiency of security code review while minimizing the risk of missing security defects. Additionally, reviewers should provide feedback on each of the detection results generated by LLMs, which can then be used to train both the difficulty classification model and LLMs for security code review.

\textbf{\textit{\black{When using a CWE list as auxiliary information for LLM in security code review, the granularity of CWE entries is crucial}}}. Our findings indicate that incorporating a CWE list into prompts significantly improves the performance of various LLMs in security defect detection, which aligns with the results of \cite{steenhoek2024comprehensive}. However, in real-world security reviews, although with experience and domain knowledge, the reviewers lack prior knowledge of all potential types of security defects that may be present in the code. It should also be noted that an overly exhaustive scope can diminish the accuracy of LLM responses. In \cite{bakhshandeh2023using}, providing a list of all 75 CWEs in their prompts decreased the accuracy of security defect detection and location by the LLMs. This drop in accuracy may be attributed to the overly lengthy and detailed list of CWEs provided, which interferes with the LLM's capability to focus effectively on the target defect, yielding adverse effects. \black{Therefore, constructing a concise yet comprehensive CWE list is necessary. In our experiments,} \textcolor{black}{due to the limited context length of the used LLMs, inputting the entire codebase during prompt design was not feasible. The lack of code context made it difficult for the LLM to fully understand the dependencies and interaction logic between files, which in turn affected its performance in identifying security defects, such as \textit{Race Condition} and \textit{Insecure Library Version}. We tested various approaches, including one-shot and few-shot prompting strategies, as well as providing the complete patchset within the prompt. These strategies required splitting the code into sections for input, which we found significantly degraded the performance of LLMs compared to the basic prompt \(\textbf{P}_b\). In order to provide as much effective information as possible within the limited context window, we designed prompts that combined multiple elements, including project metadata, CWEs, code diffs, and the commit messages. We also standardized the output format to help the model better understand the task and introduced CoT reasoning to guide the model in inferring missing context. Ultimately, we identified and selected four prompts that outperformed others, which, together with the basic prompt, were selected as templates for our formal experiments. \black{The results in RQ1.3 demonstrate that for \black{GPT-4 (ChatGPT)}, including a CWE list in the prompt significantly enhances LLM performance, while for DeepSeek-R1, the impact of the provided CWE list on model performance depends on the specific programming language. We therefore emphasize the need to explore CWE entries of different granularities for various types of security defects to determine the optimal CWE list.} As the context window of LLMs continues to expand and prompt engineering techniques evolve, we also plan to explore additional strategies in the future, such as incorporating RAG for external information, employing automated prompt optimization to dynamically adjust prompts, \black{and leveraging the agent architecture to dynamically invoke contextual information during reasoning,} thereby further improving the capability of LLMs in security defect detection.}

\textbf{\textit{With the development of LLMs, while expanding its context window, it is also necessary to ensure that LLMs can accurately capture details from long inputs.}} The emergence of LLMs that support large amounts of tokens has alleviated the challenge of inputting code context information in applying LLMs to code-related tasks. However, as evidenced by the results of RQ3, when dealing with code that contains more tokens, it is more difficult for LLMs to identify target security defects. This is because many security defects are introduced by small code snippets, and longer inputs may distract LLMs from overlooking certain details, thereby failing to detect security defects introduced by small code snippets. Additionally, according to RQ1, \black{for general-purpose LLMs,} despite GPT-4 Turbo having a context of 128K tokens, its performance in detecting security defects is noticeably inferior to that of smaller-context LLMs like GPT-4 \black{(via both API and ChatGPT)} and \textit{Llama 2}. \black{While for the reasoning-optimized LLM, although DeepSeek-R1 also has a 128k context window and is the best-performing model in this study, its rate of \textit{Incorrect fact} reaches as high as 31.81\%. Specifically, DeepSeek-R1 often presented problems with mismatched code line numbers and code snippets in its responses, which significantly undermines the understandability of its analysis. Therefore, while expanding the LLMs' context window, it is also necessary to guarantee the model’s precise memory of detailed information such as key code segments, to enhance the accuracy of security defect detection and the clarity of defect descriptions.}

\textbf{\textit{A key challenge in applying LLMs to security code review is ensuring comprehensive detection of security defects while maintaining consistent results.}} \black{The results from RQ1.4 indicate that the two LLMs, \black{i.e., DeepSeek-R1 and GPT-4 (ChatGPT),} with stronger capabilities to detect security defects tend to exhibit poorer consistency in their responses. This may be because LLMs with better performance are more capable of reasoning from multiple perspectives. The inherent randomness of LLM leads to variations in reasoning paths, resulting in inconsistent outcomes. In comparison, the thought of less capable models is relatively shallow, and their strong consistency in responses may stem from failing to deeply analyze the code. The security code review task requires models to explore as many reasoning paths as possible in order to maximize the recall of security defects, making the inconsistency of responses inevitable. In real-world application of LLMs for security code review, both the accuracy and stability of detection results are crucial.} However, current studies related to LLMs in vulnerability detection have largely overlooked the importance of the stability of results. \black{Therefore, for researchers, exploring approaches to improve the detection sensitivity while maintaining result stability, such as through more precise contextual information or voting mechanisms, is a valuable direction for future improvement.}

\black{\textit{\textbf{In practical applications, inference-time optimization adapted to downstream tasks can significantly affect both the performance and consistency of underlying LLMs.}} In RQ1, this study evaluated the capability of GPT-4 via both the ChatGPT platform and the OpenAI API in security defect detection. We found that GPT-4 (ChatGPT) significantly outperformed GPT-4 (API). Moreover, the former exhibited greater variability in outputs across multiple calls with the same input. Using the TikToken~\citep{tiktoken} tokenizer, we further calculated the average output length of GPT-4 via ChatGPT and API, which are 454 and 94 tokens, respectively. This indicates that GPT-4 via ChatGPT tends to generate more verbose and detailed responses compared to that accessed via the API. These discrepancies may be partially due to black-box optimization strategies within the ChatGPT platform, such as hyperparameter tuning, system prompts, context management mechanisms and output post-processing. As a result, compared with API-based access, ChatGPT-based access demonstrates notable improvements in the diversity of generated outputs, the level of detail in expression, and the coverage of security defect detection.}

\section{Threats to Validity} \label{sec:threat}
\textbf{Internal Validity} refers to whether an observed association between two variables can be attributed to a causal link between them. One of the internal validity threats could be \textbf{data leakage}. The data used in our study was obtained from two large-scale open-source projects, which may overlap with the data used for training LLMs, a common problem with LLM-based studies. \black{Since the dataset we used ~\citep{yu2023security} predates knowledge cutoff dates of LLMs we selected}, we cannot avoid the overlap between the dataset we used and the training data of LLMs by filtering the data for a specific period of time. Moreover, according to \cite{cao2024concerned}, contaminated data does not always affect the results, and the model sometimes even performs better after the model's cutoff date. Another threat is \textbf{the design of prompts}. We design our prompt according to best practices to evoke the best model performance. We selected words that are as unbiased, non-suggestive, and unambiguous as possible. We iteratively revised the prompts to avoid individual words from impacting the responses of LLMs during our preliminary exploration. In addition, we only included code from the single vulnerable file rather than the whole patchset in the prompt, which may affect the performance of LLMs due to \textbf{the lack of contextual information}. \black{To explore approaches for context augmentation, we designed a prompt \(\textbf{P}_{cot}\), in which LLMs are instructed to reason step-by-step—generating code context according to the commit message before detecting security defects. Since multiple factors in \(\textbf{P}_{cot}\) could influence model performance, we also designed \(\textbf{P}_{cot-guardrail}\) as a control prompt that includes only the commit message and CoT instruction, in order to isolate the effect of the fabricated code context.}

\textbf{External Validity} concerns whether a causal link generalizes across contexts. In this study, the experiments were conducted on 7 LLMs based on 4 open-source projects across programming languages including Python and C/C++. We acknowledge that the dataset we utilized may not fully represent industrial projects across various programming languages, and the performance of the selected LLMs may not represent the capabilities of all existing LLMs. We implemented the following strategies to ensure the generalizability of the experimental results. For \textbf{model diversity}, our experiments incorporated both open-source and proprietary LLMs of varying parameter scales to encompass a diverse range of popular \black{general-purpose and reasoning-optimized} LLMs. For \textbf{dataset construction}, we collected code review data from real-world open-source projects in the OpenStack and Qt communities. These projects differ in architectures, programming languages, and community structures, providing a comprehensive and valuable dataset for empirical analysis.

\textbf{Construct Validity} assesses whether a specific set of metrics corresponds to what they are intended to measure. To measure the performance of LLMs, we classified the natural language responses of LLMs into four predefined categories, with human reviews of security defects in code as the ground truth. There is a potential \textbf{natural language understanding bias} which affects classification accuracy. To reduce the bias, we established granular definitions for each category in three dimensions: defect type, location, and description, supplemented by examples for each category to ensure that the metrics are accurate, appropriate, and consistent throughout the evaluation process. In addition, the data labeling in RQ1 and the analysis of quality problems in RQ2 were all carried out manually. In order to further reduce the impact of \textbf{subjective judgments}, we employed open coding and constant comparison~\citep{glaser1965constant}. In the data labeling employed for RQ1, the first and third authors conducted a pilot data labeling and completed all the labeling work after reaching a consensus on the criteria through discussion. In the data analysis of RQ2, the first and seventh authors, conducted two rounds of pilot data analysis to determine the definition of each quality problem and achieved a high level of consistency in the results. All discrepancies in the manual work of our study were reviewed and resolved through negotiation and discussion among at least three authors, thus ensuring the reliability of the findings.   

\textbf{Conclusion Validity} refers to whether the conclusions drawn about relationships in the data are statistically sound and reasonable. Due to the \textbf{non-determinism of LLMs}, the same input may yield different outputs across different runs. Therefore, a single round of experiments is insufficient to accurately reflect the performance of LLMs. To enhance the reliability and representativeness of our evaluation, we repeated the RQ1 experiment three times. For RQ2, we calculated the average frequency of each problem across these three experiments. For RQ3, we combined the responses from the three experiments to fit the model. We also assessed their non-determinism as part of the overall performance. Specifically, in RQ1.4, we calculated the consistency of LLMs' responses across three runs to compare the non-determinism of different LLMs under different prompts.

\section{Related Work} 
\label{sec:related-work}

\subsection{Security Defect Detection in Code Review}
\textcolor{black}{Several studies have analyzed security defect detection during code reviews. \cite{paul2021security} built a dataset from the Chromium OS project, consisting of 516 code reviews that successfully identified security defects and 374 code reviews that missed security defects. The study analyzed 18 code review attributes, including review duration, reviewer experience, and author experience, to explore factors influencing the identification of security defects. It was found that the duration of code reviews and the frequency of peer reviews positively affected the detection of security defects, while the number of directories under review and the number of prior commits had a negative impact. \cite{alfadel2023empirical} examined pull requests (PR) from 10 popular JavaScript GitHub projects within the npm ecosystem, then analyzed the frequency, categories, and handling of security defects identified during code reviews. Their work indicated that only a small fraction of PRs raise security problems, and most of the identified security problems were resolved during the code review.}

\subsection{Code Characteristics of Security Defects}

Security defects mainly stem from improper system design or low-quality code, caused by a lack of security expertise or violations of secure coding practices during implementation. Many studies have focused on analyzing code for vulnerabilities and security defects in hopes of helping developers understand how security problems originate in a system, so that defects are detected early.

\cite{alves2016software} constructed a security defect dataset based on five widely used projects and 27 software metrics, such as lines of code, cyclomatic complexity and knots in the control flow graph, among others. This study demonstrated that most of these metrics are effective in distinguishing vulnerable from non-vulnerable code, thereby providing valuable support for security practitioners. \cite{meneely2013patch} conducted quantitative and qualitative analyses on 124 Vulnerability-Contributing Commits (VCC) in the Apache HTTP Server to identify characteristics of code changes with security defects. The findings suggest that longer lines of code and less experienced committers are more likely to introduce security defects. \cite{bosu2014identifying} identified 413 VCCs from code reviews in 10 open-source projects and reported consistent observations with \cite{meneely2013patch} that longer lines of code and less experienced committers are more likely to introduce security defects. Additionally, they indicated that modified code files are more likely to contain security defects than new files.

\subsection{Automated Tools of Security Defect Detection}
 
Various tools have been developed to automatically detect security defects. Since our study focuses on using LLMs to assist reviewers in identifying security defects during code review,  we limit our coverage to studies related to security defect detection, excluding those studies on security defect remediation or program repair.

\cite{charoenwet2024empirical} conducted an empirical study on the application of static application security testing tools (SASTs) in security code review and found that a single SAST tool could produce warnings for 52\% of VCCs. By combining several tools, warnings could be produced for 78\% of the VCCs. However, at least 76\% of these warnings are irrelevant to the vulnerability in VCCs, highlighting the challenges that SAST tools face in providing detailed information about the vulnerability during security code review. Inspired by this work, \textbf{our work} investigated the application of LLMs in security code review, aiming to overcome the limitations of SAST tools to some extent.

Several studies have explored the performance of LLMs in detecting security defects. The majority of these studies conducted a coarse-grained assessment of LLMs in binary classification tasks that determine whether the code contains any security defects without specifying the corresponding defect type. \cite{zhou2024largelanguage} utilized CWE, project information, and a few-shot learning approach to design prompts and evaluated the capability of two popular LLMs, GPT-3.5 and GPT-4, in binary judgment under various prompts. They found that GPT-3.5 achieved competitive performance with the prior state-of-the-art detection approaches, while GPT-4 consistently outperformed the state-of-the-art. \cite{purba2023software} compared the performance of LLMs with popular static analysis tools in binary detection. Their work demonstrates that although LLMs achieved a good recall rate, their false positive rate was significantly higher than that of static analysis tools. \black{~\cite{jensen2024software} evaluated the performance of three proprietary OpenAI LLMs and four small-scale models with fewer than 13B parameters to identify vulnerabilities in the SecurityEval dataset~\citep{siddiq2022securityeval}. They adopted zero-shot and CoT prompting and found that the best‑performing model - GPT‑4 - achieved an accuracy of 95.6\% in distinguishing between vulnerable and patched versions of code. Similarly, ~\cite{steenhoek2024err} assessed the capabilities of 14 LLMs on a real-world and function-level vulnerability dataset, SVEN, across seven different prompt strategies. They explored various approaches, including prompt optimization, scaling model size, increasing pre-training data, fine-tuning, and incorporating domain knowledge into prompts, yet none led to significant improvements in the performance of models in the binary vulnerability detection task.} Unlike previous studies, \textbf{our work} provides a more fine-grained evaluation of the capability of LLMs in security code reviews by including additional insights such as defect localization, description, and suggestions for defect remediation.

Previous studies also investigated whether LLMs could provide more information in security defect detection. \cite{bakhshandeh2023using} provided LLMs with a list of CWEs and the inspection results of static analysis tools on the prompts, requesting the LLMs to generate specific CWE names along with the corresponding code line numbers. Their study shows that LLMs effectively reduce false positives of static tools and demonstrate competitive or even superior performance in defect localization compared to these tools. \cite{yin2024multitask} evaluated the performance of LLMs across multiple tasks in security analysis, including defect detection, localization, and description, and found that the existing state-of-the-art approaches, such as LineVul~\citep{fu2022linevul} and SVulD~\citep{ni2023distinguishing}, generally outperformed LLMs. Specifically, LLMs exhibited limited overall accuracy and varied performance across different CWE types in defect localization and description tasks. \black{~\cite{khare2025understanding} comprehensively evaluated 16 pre-trained LLMs for detecting security defects on a function-level vulnerability dataset of 5,000 samples. Their study systematically compared model performance across three distinct prompting strategies and found that involving step-by-step analysis significantly improved LLM performance. This work also performed comparison of LLMs against leading static analysis tools and deep learning-based approaches, finding that LLMs reported higher detection accuracies than them.} These studies have all been limited to quantitatively evaluating LLM performance. In comparison, \textbf{our work} performed an in-depth analysis of the quality problems in LLM-generated responses and factors affecting LLM performance to understand the potential and challenges LLMs face in security code review.

\black{Fine-tuning has also been employed to improve the capability of LLMs in security defect detection. ~\cite{mechri2025secureqwen} fine-tuned CodeQwen1.5-7B on a large Python dataset comprising both real-world code and synthetic code generated by LLMs, in order to guide the model to capture and classify security defects in code accurately. The model ultimately achieved high accuracy, with F1 scores ranging from 84\% to 99\%. Although fine-tuning can effectively improve the accuracy of model, fine-tuned LLMs often exhibit poor generalization.  In practical application, adjustments to the model parameters and subsequent retraining are usually required. Consequently, some studies have shifted their focus toward approaches that avoid altering the core model, such as leveraging prompt engineering or incorporating external information to further enhance the precision of LLMs in security defect detection.}
\cite{yang2024dlap} introduced a framework utilizing DL models, integrating context learning and CoT to augment prompts in security defect detection. Their framework yielded superior performance compared to state-of-the-art prompting techniques. \cite{nong2024chain} proposed a vulnerability-semantics-guided prompting (VSP) approach that combines CoT with various auxiliary information to evaluate the performance of LLMs in identifying security defects (including their classification and resolution). The research concerns and employed datasets of the above two studies~\citep{yang2024dlap,nong2024chain} differ significantly from ours. These studies focus on designing benchmarks to enhance LLM in vulnerability detection, utilizing synthetic datasets and CVE datasets composed of publicly disclosed vulnerabilities to validate their improvements. Compared with them, \textbf{our work} focuses on the empirical evaluation of LLMs in the context of security code review, employing a dataset of security defects collected from code reviews of real-world software projects. \black{To further improve the accuracy of LLMs in security defect detection, some studies have adopted RAG to retrieve external knowledge and enhance the input prompts. \cite{du2024vul} constructed multi-dimensional representations to retrieve knowledge most relevant to the target code from historical vulnerabilities and their corresponding fixes. This framework improved the accuracy of LLMs in distinguishing vulnerable code from patched code by 16\%–24\%. \cite{lu2024grace} incorporated graph-structured information from the code and identified highly related code samples as external guidance into the prompt. Using this method, LLMs outperformed state-of-the-art deep learning-based tools in detecting vulnerabilities. With the emergency of reasoning-optimized LLMs, LLM-based agents have also shown promising applications in security defect detection. Inspired by courtroom, \cite{widyasari2025let} constructed a framework consisting of four distinct agent roles, which are security researcher, code author, moderator, and review board. Their framework employs multiple rounds of interaction to arrive at a final determination on security defects in the code, which substantially surpassed the performance of approaches based on CoT and fine-tuning. \cite{tsai2025leveraging} incorporated RAG into the multi-turn multi-role interaction agent framework, achieving a 29.9\% improvement in F1 score compared to~\cite{widyasari2025let}.} It is worth noting that the studies mentioned above focus on general vulnerability detection by LLMs, whereas \textbf{our work} targets the specific area of security code review by LLMs. Inspired by these related works, we aim to leverage LLMs' potential in security analysis to identify the obstacles, application strategies, and improvement directions for integrating LLMs into security code review.
\vspace*{-1em}
\section{Conclusions and Future Work} \label{sec:conclusions}
In our study, we comprehensively analyzed the practicality of current popular LLMs in the context of security code review. We investigated the capabilities of \black{7} LLMs in detecting security defects similar to those identified by human reviewers during real-world code reviews. \black{For the two top-performing LLMs, each paired with their respective optimal prompts, we conducted a linguistic analysis of the quality problems in their generated responses. Furthermore, we performed separate regression analyses to explore the factors influencing the performance of each LLM-prompt combination.}

Our main findings are: (1) In security code review tasks, LLMs demonstrate stronger capabilities than state-of-the-art static analysis tools. Among these LLMs, the reasoning-optimized model outperforms established general-purpose models. However, their practical application remains limited. 
(2) DeepSeek-R1 is the top-performing LLM in security code review, followed by \black{GPT-4 (ChatGPT)}. DeepSeek-R1 performs best when the prompt includes the commit message and CoT guidance, whereas \black{GPT-4 (ChatGPT)} performs best when provided with a CWE list in the prompt. 
(3) DeepSeek-R1 and \black{GPT-4 (ChatGPT)} are more effective at detecting security defects in files with fewer tokens and with security-related code annotations. \black{DeepSeek-R1 tends to perform better on detecting certain types of security defects in code with higher complexity.}
(4) The most prevalent quality problem for both DeepSeek-R1 and \black{GPT-4 (ChatGPT)} is verbosity. In comparison, \black{GPT-4 (ChatGPT)} tends to produce vague expressions and demonstrates poorer instruction-following ability, while DeepSeek-R1 is more prone to errors in recalling code details.

\black{In future work, we plan to explore several directions to further enhance the capabilities of LLMs in security code review: (1) Improving the sensitivity to security defects of LLMs, by investigating the optimal granularity of the CWE list provided in the prompt and incorporating external knowledge or project-specific context to ensure comprehensive consideration of frequent security defect types; (2) Enhancing contextual reasoning by developing approaches to provide more precise code context to inform the analysis procedure; (3) Increasing the stability of the LLMs' outputs via techniques such as self-refinement and voting mechanisms, to generate more consistent and reliable results.}


\section*{Data availability}
We have shared the link to our dataset in the reference~\citep{replpack}. 

\section*{Acknowledgments}
This work is supported by the National Natural Science Foundation of China under Grant No. 92582203 and the Major Science and Technology Project of Hubei Province under Grant No. 2024BAA008. The numerical calculations in this paper have been done on the supercomputing system in the Supercomputing Center of Wuhan University.

\printcredits

\bibliographystyle{cas-model2-names}
\bibliography{references}
\balance
\end{sloppypar}
\end{document}